\documentclass[english,prc,showpacs,reprint,amsmath,amssymb,superscriptaddress,floatfix]{revtex4-1}
\usepackage{dcolumn}
\usepackage{color}
\usepackage[dvips]{graphicx}
\usepackage{amsmath}
\usepackage{multirow}
\usepackage{bm}
\usepackage[T1]{fontenc}
\usepackage[latin9]{inputenc}
\usepackage{color}
\usepackage{amssymb}
\usepackage{float}

\providecommand{\tabularnewline}{\\}

\newcommand\T{\rule{0pt}{3.1ex}}

\usepackage{array}\usepackage{babel}

\providecommand{\tabularnewline}{\\}

\usepackage{babel}
\begin{document}

\title{Nuclear matrix elements for double-$\beta$ decay}

\author{J.\ Barea}
\email{jbarea@udec.cl}
\affiliation{Departamento de F\'{i}sica, Universidad de Concepci\'{o}n,
 Casilla 160-C, Concepci\'{o}n, Chile}

\author{J. Kotila}
\email{jenni.kotila@yale.edu}
\affiliation{Center for Theoretical Physics, Sloane Physics Laboratory,
 Yale University, New Haven, Connecticut 06520-8120, USA}

\author{F.\ Iachello}
\email{francesco.iachello@yale.edu}
\affiliation{Center for Theoretical Physics, Sloane Physics Laboratory,
 Yale University, New Haven, CT 06520-8120, USA}

\begin{abstract}
\textbf{Background}: Direct determination of the neutrino mass through double-$\beta$ decay is at the present time one of the most important areas
of experimental and theoretical research in nuclear and particle physics.\\
\textbf{Purpose}: We calculate nuclear matrix elements for the extraction of the
average neutrino mass in neutrinoless double-$\beta$ decay. Methods:
The microscopic interacting boson model (IBM-2) is used.\\ 
\textbf{Results}:
Nuclear matrix elements in the closure approximation are calculated
for $^{48}$Ca, $^{76}$Ge, $^{82}$Se, $^{96}$Zr, $^{100}$Mo, $^{110}$Pd,
$^{116}$Cd, $^{124}$Sn, $^{128}$Te, $^{130}$Te, $^{148}$Nd, $^{150}$Nd,
$^{154}$Sm, $^{160}$Gd, and $^{198}$Pt decay.\\
 \textbf{Conclusions}: Realistic
predictions for the expected half-lives in neutrinoless double-$\beta$
decay with light and heavy neutrino exchange in terms of neutrino
masses are made and limits are set from current experiments. 
\end{abstract}


\maketitle

\section{INTRODUCTION}

\begin{figure}[h]
\includegraphics[width=8.6cm]{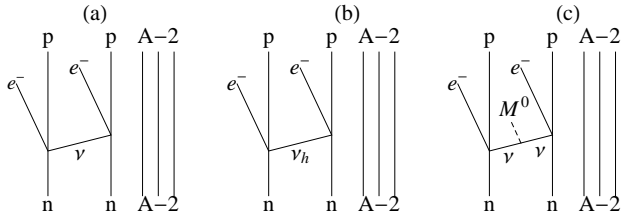} 
\caption{\label{fig1} Neutrinoless double-$\beta$ decay mechanism for (a) light neutrino exchange, (b) heavy neutrino exchange, and (c) Majoron emission.}
\end{figure}

In recent years, the possibility of a direct measurement of the average
neutrino mass in neutrinoless double-$\beta$ decay has attracted considerable
attention. Three scenarios have been considered \cite{doi,tomoda,simkovic}, shown in Fig.~\ref{fig1}. 
After the discovery of neutrino oscillations \cite{fukuda,ahmad,eguchi}, attention has been focused on the first scenario (a).
In very recent years, the second scenario (b) has again attracted
attention \cite{tello}. For all three processes (0$\nu\beta\beta$,
0$\nu_{h}\beta\beta$, and 0$\nu\beta\beta M$), the half-life can
be factorized as 
\begin{equation}
\lbrack\tau_{1/2}^{0\nu}]^{-1}=G_{0\nu}\left\vert M_{0\nu}\right\vert ^{2}\left\vert f(m_{i},U_{ei})\right\vert ^{2},
\end{equation}
where $G_{0\nu}$ is a phase-space factor, $M_{0\nu}$ is the nuclear
matrix element, and $f(m_{i},U_{ei})$ contains physics beyond the
standard model through the masses $m_{i}$ and mixing matrix elements
$U_{ei}$ of neutrino species.

\begin{figure}[h]
\begin{center}
\includegraphics[width=3.5cm]{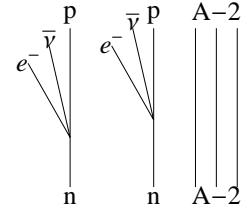} 
\end{center}
\caption{\label{fig2} Double-$\beta$ decay mechanism with the emission of $2\bar{\nu}$.}
\end{figure}

In addition to the neutrinoless modes, there is also the process allowed
by the standard model, 2$\nu\beta\beta$, depicted in Fig.~\ref{fig2}. 
For this process, the half-life can be, to a good approximation, factorized
in the form 
\begin{equation}
\left[\tau_{1/2}^{2\nu}\right]^{-1}=G_{2\nu}\left\vert M_{2\nu}\right\vert ^{2}.
\end{equation}
(The factorization here is not exact and conditions under which
it can be done are discussed in Ref.~\cite{kotila} and Sec. III).

The processes depicted in Figs.~\ref{fig1} and \ref{fig2} are of the type 
\begin{equation}
\left(A,Z\right)\rightarrow\left(A,Z+2\right)+2e^{-}+\text{anything.}
\end{equation}
 In very recent years, interest in the processes 
\begin{equation}
\left(A,Z\right)\rightarrow(A,Z-2)+2e^{+}+\text{anything}
\end{equation}
 has also arisen. In this case there are also the competing modes
in which either one or two electrons are captured from the electron
cloud (0$\nu\beta$EC, 2$\nu\beta$EC, and 2$\nu$ECEC). Also for these
modes, the half-life can be factorized (either exactly or approximately)
into the product of a phase-space factor and a nuclear matrix element,
which then are the crucial ingredients of any double-$\beta$ decay calculation.

Recently, we have initiated a program for the systematic evaluation
of both quantities. The evaluation of the phase-space factors (PSFs)
for 0$\nu\beta^{-}\beta^{-}$ and 2$\nu\beta^{-}\beta^{-}$ has been
reported in \cite{kotila} and that for 0$\nu\beta^{+}\beta^{+}$,
2$\nu\beta^{+}\beta^{+}$, 0$\nu\beta^{+}$EC, 2$\nu\beta^{+}$EC, and
2$\nu$ECEC is in preparation \cite{kotila2}. The main difference betveen this  new calculation of PSFs and older standard  approximations is a few percent for light nuclei ($Z=20$), about 30\% for Nd ($Z=60$), and a rather large 90\% for U ($Z=92$), the correction increasing as a power of $Z\alpha$.  In this article, we
concentrate on nuclear matrix elements $M_{0\nu}$ and $M_{2\nu}$
for 0$\nu\beta^{-}\beta^{-}$ and 2$\nu\beta^{-}\beta^{-}$. Calculations
of the nuclear matrix elements for positron emission have also been
completed and will be reported in a subsequent paper \cite{barea2}.
Nuclear matrix elements have been evaluated in a variety of models, most notably the quasiparticle random phase approximation (QRPA) 
and the interacting shell model (ISM). Results up to 1998 are reviewed in Refs. \cite{suhonen1998} and \cite{faessler1998}. In 1999 a new formulation of $0\nu\beta\beta$ was introduced \cite{simkovic} and calculations within the QRPA \cite{simkovic1} and the ISM \cite{poves} were performed, as well as within other models, as discussed in the following Sec. II C. In 2009, we developed
\cite{barea} a new method to evaluate nuclear matrix elements for
double-$\beta$ decay within the framework of the microscopic interacting
boson model (IBM-2). The advantage of this method is that it can be
used in any nucleus and thus all nuclei of interest in both  $\beta^{-}\beta^{-}$
and $\beta^{+}\beta^{+}$ decay can be calculated within the same model.

The calculation of the nuclear matrix elements is done in the \textit{closure}
approximation. This approximation is good for 0$\nu\beta\beta$ decay,
since the average neutrino momentum is of the order of 100~MeV/$c$. It
is, in principle, not good for 2$\nu\beta\beta$, since the average
neutrino momentum is of the order of few MeV/$c$. However, formally
the approximation is still valid if one appropriately defines the
closure energy. The advantage of the closure approximation is that
all calculations for the processes depicted in Figs.~\ref{fig1} and \ref{fig2} can be
done simultaneously, by changing the so-called neutrino potential,
as discussed in the sections below, thus eliminating systematic (and
accidental) errors in the calculation, especially in the ratio of
matrix elements for different processes.

\begin{ruledtabular}
\begin{center}
\begin{table}[h]
\caption{\label{table1}Double-$\beta$ decays considered in this article, their $Q$-values, and their isotopic abundances.}
\begin{tabular}{lll}
$\beta^{-}\beta^{-}$ transition &$Q_{\beta\beta}$(keV)&$P$(\%) \\
\hline
\T
$_{20}^{48}$Ca$_{28}\rightarrow _{22}^{48}$Ti$_{26}$ &$4272.26 \pm 4.04$  &$0.187\pm 0.021$ \\
$_{32}^{76}$Ge$_{44}\rightarrow _{34}^{76}$Se$_{42}$ &$2039.061 \pm 0.007$  &$7.73\pm 0.12$\\
$_{34}^{82}$Se$_{48}\rightarrow _{36}^{82}$Kr$_{46}$ &$2995.12 \pm 2.01$  &$8.73\pm 0.22$ \\
$_{40}^{96}$Zr$_{56}\rightarrow _{42}^{96}$Mo$_{54}$ &$3350.37 \pm 2.89$  &$2.80\pm 0.09$ \\
$_{42}^{100}$Mo$_{58}\rightarrow _{44}^{100}$Ru$_{56}$ &$3034.40 \pm 0.17$  &$9.82\pm 0.31$ \\
$_{46}^{110}$Pd$_{64}\rightarrow _{48}^{110}$Cd$_{62}$ &$2017.85 \pm 0.64$  &$11.72\pm 0.09$ \\
$_{48}^{116}$Cd$_{68}\rightarrow _{50}^{116}$Sn$_{66}$ &$2813.50 \pm 0.13$  &$7.49\pm 0.18$ \\
$_{50}^{124}$Sn$_{74}\rightarrow _{52}^{124}$Te$_{72}$ &$2286.97 \pm 1.53$  &$5.79\pm 0.05$ \\
$_{52}^{128}$Te$_{76}\rightarrow _{54}^{128}$Xe$_{74}$ &$865.87 \pm 1.31$  &$31.74\pm 0.08$ \\
$_{52}^{130}$Te$_{78}\rightarrow _{54}^{130}$Xe$_{76}$ &$2526.97 \pm 0.23$  &$34.08\pm 0.62$ \\
$_{54}^{136}$Xe$_{82}\rightarrow _{56}^{136}$Ba$_{80}$ &$2457.83 \pm 0.37$  &$8.8573\pm 0.0044$ \\
$_{60}^{148}$Nd$_{88}\rightarrow _{62}^{148}$Sm$_{86}$ &$1928.75 \pm 1.92$  &$5.756\pm 0.021$ \\
$_{60}^{150}$Nd$_{90}\rightarrow _{62}^{150}$Sm$_{88}$ &$3371.38 \pm 0.20$  &$5.638\pm 0.028$  \\
$_{62}^{154}$Sm$_{92}\rightarrow _{64}^{154}$Gd$_{90}$ &$1251.03 \pm 1.25$  &$22.75\pm 0.29$ \\
$_{64}^{160}$Gd$_{96}\rightarrow _{66}^{160}$Dy$_{94}$ &$1729.69 \pm 1.26$  &$21.86\pm 0.19$ 	  \\
$_{78}^{198}$Pt$_{120}\rightarrow _{80}^{198}$Hg$_{118}$ &$1047.17 \pm 3.11$  &$7.36\pm 0.13$  \\
\end{tabular}
\end{table}
\end{center}
\end{ruledtabular}

In this article, we report the results of our calculations for the
nuclei listed in Table~\ref{table1}. A selected number of decays wer considered in \cite{barea} and preliminary
results were presented in \cite{iac-barea-otranto,iac-barea-medex11}.
Here we report the complete list of results divided into 0$\nu\beta\beta$
(light neutrino exchange) and 0$\nu_{h}\beta\beta$ (heavy neutrino
exchange), Sec. II, and 2$\nu\beta\beta$, Sec. III. By using these
results we also set some limits on the mass of light $\left\langle m_{\nu}\right\rangle $
and heavy $\left\langle m_{\nu_{h}}\right\rangle $ neutrinos.

\section{NEUTRINOLESS\ DOUBLE\-$\beta$\ DECAY (0$\nu\beta\beta$)}
\label{sect0nu}

\subsection{Transition operator}

The theory of 0$\nu\beta\beta$ decay was first formulated by Furry
\cite{furry} and further developed by Primakoff and Rosen \cite{primakoff},
Molina and Pascual \cite{molina}, Doi \textit{et al.} \cite{doi},
Haxton and Stephenson \cite{haxton}, and, more recently, by Tomoda
\cite{tomoda} and \v{S}imkovic \textit{et al.} \cite{simkovic}. All
these formulations often differ by factors of 2, by the number of
terms retained in the nonrelativistic expansion of the current and
by their contribution. In order to have a standard set of calculations
to be compared with the QRPA and the ISM, we adopt in this article the formulation
of \v{S}imkovic \textit{et al.} \cite{simkovic}. The transition operator
in momentum space, $p=\left\vert \vec{q}\right\vert $, is written
as 
\begin{equation}
T(p)=H(p)f(m_{i},U_{ei})
\end{equation}
where for light neutrino exchange 
\begin{equation}
f\left(m_{i},U_{ei}\right)=\frac{\left\langle m_{\nu}\right\rangle }{m_{e}}\text{, \ \ \ }\left\langle m_{\nu}\right\rangle =\sum_{k=light}\left(U_{ek}\right)^{2}m_{k}\text{,}
\end{equation}
while for heavy neutrino exchange 
\begin{equation}
\begin{split}
f(m_{i},U_{ei})=m_{p}\left\langle m_{\nu_{h}}^{-1}\right\rangle,
 \\
\left\langle m_{\nu_{h}}^{-1}\right\rangle =\sum_{k=heavy}\left(U_{ek_h}\right)^{2}\frac{1}{m_{k_h}}.
\end{split}
\end{equation}
The (two-body) operator $H(p)$ can be written as 
\begin{equation}
\begin{split}
H(p)=\sum_{n,n^{\prime}}\tau_{n}^{\dag}\tau_{n^{\prime}}^{\dag}\left[-h^{F}(p)+h^{GT}(p)\vec{\sigma}_{n}\cdot\vec{\sigma}_{n^{\prime}} \right.
\\
\left. +h^{T}(p)S_{nn^{\prime}}^{p}\right],
\end{split}
\end{equation}
with the tensor operator defined as 
\begin{equation}
S_{nn^{\prime}}^{p}=3\left[\left(\vec{\sigma}_{n}\cdot\hat{p}\right)\left(\vec{\sigma}_{n^{\prime}}\cdot\hat{p}\right)\right]-\vec{\sigma}_{n}\cdot\vec{\sigma}_{n^{\prime}}.
\end{equation}
The Fermi (F), Gamow-Teller (GT), and tensor (T) contributions are
given by 
\begin{equation}
\begin{split}
h^{F}(p) & =  h_{VV}^{F}(p)\\
h^{GT}(p) & =  h_{AA}^{GT}(p)+h_{AP}^{GT}(p)+h_{PP}^{GT}(p)+h_{MM}^{GT}(p)\\
h^{T}(p) & =  h_{AP}^{T}(p)+h_{PP}^{T}(p)+h_{MM}^{T}(p).
\end{split}
\end{equation}
The terms AP, PP, and MM are higher order corrections (HOC) arising from
weak magnetism (M) and induced pseudoscalar terms (P) in the weak
nucleon current. The terms $h^{F,GT,T}(p)$
can be further factorized as 
\begin{equation}
h^{F,GT,T}(p)=v(p)\tilde{h}^{F,GT,T}(p)
\end{equation}
where $v(p)$ is called the neutrino potential
and are the $\tilde{h}^{F,GT,T}(p)$ the form factors.
A list of form factors is given in Ref.~\cite{simkovic} and recast
in the form used by us in Table~\ref{table2}. 
In this table, the finite nucleon size (FNS) is taken into account
by taking the coupling constants $g_{V}$ and $g_{A}$ as momentum dependent
\begin{equation}
\label{ffeq}
\begin{split}
g_{V}(p^{2}) & =  g_{V}\frac{1}{\left(1+\frac{p^{2}}{M_{V}^{2}}\right)^{2}}, \\
g_{A}(p^{2}) & =  g_{A}\frac{1}{\left(1+\frac{p^{2}}{M_{A}^{2}}\right)^{2}}.
\end{split}
\end{equation}
The value of $M_{V}$ is well fixed by the electromagnetic form factor
of the nucleon, $M_{V}^{2}=0.71$(GeV/$c^{2}$)$^{2}$ \cite{nucleonem}
and $g_{V}=1$ by the hypothesis of conserved vector current (CVC). The value of $M_{A}$ is estimated to be $M_{A}=1.09$(GeV/$c^{2}$)
\cite{nucleonaxial} and $g_{A}=1.269$ \cite{experiment}.

\begin{ruledtabular}
\begin{center}
\begin{table}[h]
\caption{\label{table2}Form factors in the formulation of \cite{simkovic} adapted to our calculation. $m_{p}$ and $m_{\pi}$ are, respectively, the proton and pion mass and $\kappa_{\beta}=3.70$ is the isovector anomalous magnetic moment of the nucleon.}
\begin{tabular}{cc}
Term  &  \ensuremath{\tilde{h}(p)}\\
\hline
\T
 \ensuremath{\tilde{h}_{VV}^{F}} &  \ensuremath{g_A^2\frac{(g_{V}^{2}/g_{A}^{2})}{\left(1+p^{2}/M_{V}^{2}\right)^{4}}}\\
\ensuremath{\tilde{h}_{AA}^{GT}} &  \ensuremath{\frac{g_{A}^{2}}{\left(1+p^{2}/M_{A}^{2}\right)^{4}}}\\
\ensuremath{\tilde{h}_{AP}^{GT}} &  \ensuremath{g_{A}^{2}\left[-\frac{2}{3}\frac{1}{\left(1+p^{2}/M_{A}^{2}\right)^{4}}\frac{p^{2}}{p^{2}+m_{\pi}^{2}}(1-\frac{m_{\pi}^{2}}{M_{A}^{2}})\right]}\\
\ensuremath{\tilde{h}_{PP}^{GT}} &  \ensuremath{g_{A}^{2}\left[\frac{1}{\sqrt{3}}\frac{1}{\left(1+p^{2}/M_{A}^{2}\right)^{2}}\frac{p^{2}}{p^{2}+m_{\pi}^{2}}\left(1-\frac{m_{\pi}^{2}}{M_{A}^{2}}\right)\right]^2}\\
\ensuremath{\tilde{h}_{MM}^{GT}} &  \ensuremath{g_{A}^{2}\left[\frac{2}{3}\frac{g_{V}^{2}}{g_{A}^{2}}\frac{1}{\left(1+p^{2}/M_{V}^{2}\right)^{4}}\frac{\kappa_{\beta}^2 p^{2}}{4m_{p}^{2}}\right]}\\
\ensuremath{\tilde{h}_{AP}^{T}} &  \ensuremath{-\tilde{h}_{AP}^{GT}}\\
\ensuremath{\tilde{h}_{PP}^{T}} &  \ensuremath{-\tilde{h}_{PP}^{GT}}\\
\ensuremath{\tilde{h}_{MM}^{T}} &  \ensuremath{\frac{1}{2}\tilde{h}_{MM}^{GT}}
 \end{tabular}\text{ }
\end{table}
\end{center}
\end{ruledtabular}

The neutrino potential $v(p)$ is given, in the closure approximation,
for light neutrino exchange by 
\begin{equation}
v(p)=\frac{2}{\pi}\frac{1}{p\left(p+\tilde{A}\right)}.\label{eq:pot_light}
\end{equation}
For heavy neutrino exchange, the neutrino potential is given by 
\begin{equation}
v(p)=\frac{2}{\pi}\frac{1}{m_{e}m_{p}}.\label{eq:pot_heavy}
\end{equation}

The contributions in momentum space, $h^{F,GT,T}(p)$, can be converted to the contributions in coordinate space, $h^{F,GT,T}(r)$, by taking the Fourier-Bessel transforms 
\begin{equation}
\begin{split}
h^{F,GT,T}(r)=&\frac{2}{\pi}\int_{0}^{\infty}j_{\lambda}(pr) \frac{1}{p\left(p+\tilde{A}\right)}\tilde{h}^{F,GT,T}(p)\\
&\times p^{2}dp,
\end{split}
\end{equation}
for light-neutrino exchange and 
\begin{equation}
\begin{split}
h^{F,GT,T}(r)=&\frac{2}{\pi}\int_{0}^{\infty}j_{\lambda}(pr)\frac{1}{m_{e}m_{p}}\tilde{h}^{F,GT,T}(p)\\
&\times p^{2}dp,
\end{split}
\end{equation}
for heavy neutrino exchange. Here $\lambda=0$ for Fermi and Gamow-Teller contributions
and $\lambda=2$ for tensor contributions.

Finally, an additional improvement is the introduction of short-range
correlations (SRC). These can be taken into account by multiplying
the potential $V(r)$ in coordinate space by a correlation function
$f(r)$ squared. The most commonly used correlation function is the
Jastrow function 
\begin{equation}\label{jastrow}
f_{J}(r)=1-ce^{-ar^{2}}(1-br^{2})
\end{equation}
with $a=1.1$~fm$^{-2}$, $b=0.68$~fm$^{-2}$ and $c=1$ for the phenomenological
Miller-Spencer parametrization \cite{miller-spencer}, and, in recent
years, the Argonne/CD-Bonn parametrizations \cite{argonnebonn} $a=1.59/1.52\mbox{~fm}^{-2}$,
$b=1.45/1.88\mbox{~fm}^{-2}$ and $c=0.92/0.46$. Since our formulation
is in momentum space, we take into account SRC by using the Fourier-Bessel
transform of $f_{J}(r)$.

In assessing the \textquotedbl{}{}goodness\textquotedbl{}{} of \v{S}imkovic's
formulation it is of interest to compare it with Tomoda's formulation.
Apart from some differences in definitions, namely the fact that Tomoda
defines the transition operator with a factor of 1/2 in front of Eq.~(8),
see Eq.~(3.31) of Ref.~\cite{tomoda}, and the tensor operator with
a factor of 1/3 in front of Eq.~(9), see Eq.~(3.54) of Ref.~\cite{tomoda}
and with a plus sign in front of the tensor operator in Eq.~(8), in
contrast with Eq.~(13) of Ref.~\cite{simkovic}, and a nuclear radius
$R=r_{0}A^{1/3}$ with $r_{0}=1.2$~fm instead of $r_{0}=1.1$~fm of
\cite{simkovic}, differences which have caused, \ however, considerable
confusion in the literature, the main difference between Tomoda's formulation
and \v{S}imkovic's formulation is that Tomoda considers more terms in $H(p)$,
nine in all, three GT terms, three F terms, one T term, one pseudoscalar (P) term,  and one recoil (R) term.
Also, except for the terms $h_{VV}(p)$ and $h_{AA}(p)$, where the
form factors and potentials coincide, all other form factors and potentials in Tomoda's formulation
are different from those in \v{S}imkovic's formulation. Although
in this article we report results in the latter formulation, we note
that we have results available for seven of the nine terms in Tomoda's
formulation, the three GT terms, the trhree F terms, and the T term. The form factors and potentials for these
seven terms are listed in Table~VIII of our Ref.~\cite{barea}. In
Table~II of the same reference, we also show that the contribution
of additional terms $\chi_{GT}^{\prime},\chi_{F}^{\prime},$ and $\chi_{T}^{\prime}$
is not negligible and thus the assessment of the \textquotedbl{}{}goodness\textquotedbl{}{}
of \v{S}imkovic's formulation must reflect this point.

\subsection{Matrix elements}

\begin{figure}[h]
\begin{center}
\includegraphics[width=8.6cm]{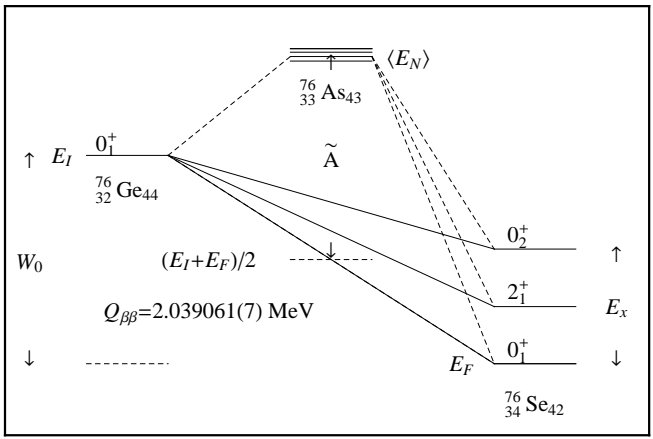} 
\end{center}
\caption{\label{fig3} The decay $_{32}^{76}$Ge$_{44}  \rightarrow _{34}^{76}$Se$_{42}$, an example of double-$\beta$ decay.}
\end{figure}

We consider the decay of a nucleus $_{Z}^{A}$X$_{N}$ into a nucleus
$_{Z+2}^{A}$Y$_{N-2}$. An example is shown in Fig.~\ref{fig3}. 
The nuclear matrix elements are those of the operator $H(p)$ of
Eq.~(8) between the ground state of the initial nucleus and the final
state with angular momentum $J_{F}$ 
\begin{equation}
M_{0\nu}\equiv\left\langle ^{A}\text{X;}0_{1}^{+}\left\vert H(p)\right\vert ^{A}\text{Y;}J_{F}\right\rangle .
\end{equation}
If the decay proceeds through an $s$-wave, with two leptons in the
final state we cannot form an angular momentum greater than one. We
therefore calculate, in this article, only $0\nu\beta\beta$ matrix elements to final
$0^{+}$ states, the ground state $0_{1}^{+}$, and the first excited
state $0_{2}^{+}$, for which in a previous article \cite{kotila}
we have calculated the phase-space factors. The form
factors in Table~\ref{table2} have a common factor of $g_{A}^{2}$, except $\ensuremath{\tilde{h}_{VV}^{F}}$.
They depend on $g_A^2$ and $g_V^2$. We write 
\begin{equation}
\begin{split}
M_{0\nu} & =  g_{A}^{2}M^{(0\nu)}, \\
M^{(0\nu)} & =  M_{GT}^{(0\nu)} -\left(\frac{g_{V}}{g_{A}}\right)^2 M_{F}^{(0\nu)}+M_{T}^{(0\nu)},
\end{split}
\end{equation}
with the ratio $g_{V}/g_{A}$ explicitly displayed in front of $M_{F}^{(0\nu)}$. The ratio $g_{V}/g_{A}$ is also implicitly contained in $M_{GT}^{(0\nu)}$ and $M_{T}^{(0\nu)}$ through the terms $\tilde{h}^{GT}_{MM}$ and $\tilde{h}^{T}_{MM}$ (see Table~\ref{table2}), and the matrix elements $M^{(0\nu)}_{GT}$, $M^{(0\nu)}_{F}$, and $M^{(0\nu)}_{T}$ are defined as
\begin{equation}
\begin{split}
M^{(0\nu)}_{GT}\equiv\left\langle ^{A}\text{X;}0_{1}^{+}\left\vert
h^{GT}(p)/g^2_A \right\vert ^{A}\text{Y;}J_{F}\right\rangle, \\ 
M^{(0\nu)}_F\equiv\left\langle ^{A}\text{X;}0_{1}^{+}\left\vert
 h^{F}(p)/g^2_V \right\vert ^{A}\text{Y;}J_{F}\right\rangle ,\\
M^{(0\nu)}_T\equiv\left\langle ^{A}\text{X;}0_{1}^{+}\left\vert 
h^{T}(p)/g^2_A \right\vert ^{A}\text{Y;}J_{F}\right\rangle .
\end{split}
\end{equation}
The reason for this separation is that the calculated single-$\beta$ decay matrix elements of the GT operator in a particular nuclear model appear to be systematically larger than those derived from the measured $ft$ values of the allowed GT transitions. The simplest way of taking into account this result is by introducing an effective $g_{A,eff}$, also sometimes written as $g_{A,eff}=qg_A$, where $q$ is a quenching factor. The quenching of $g_A$ will be discussed in Sec.~\ref{sect2nu}. Here we report results of the calculation of $M^{(0\nu)}$ in IBM-2 with the free values $g_V=1$ and $g_A=1.269$. These form the baseline for any discussion of the nuclear matrix elements (NME) in $0\nu\beta\beta$.

\begin{figure*}[cbt!]
\begin{center}
\includegraphics[width=12.cm]{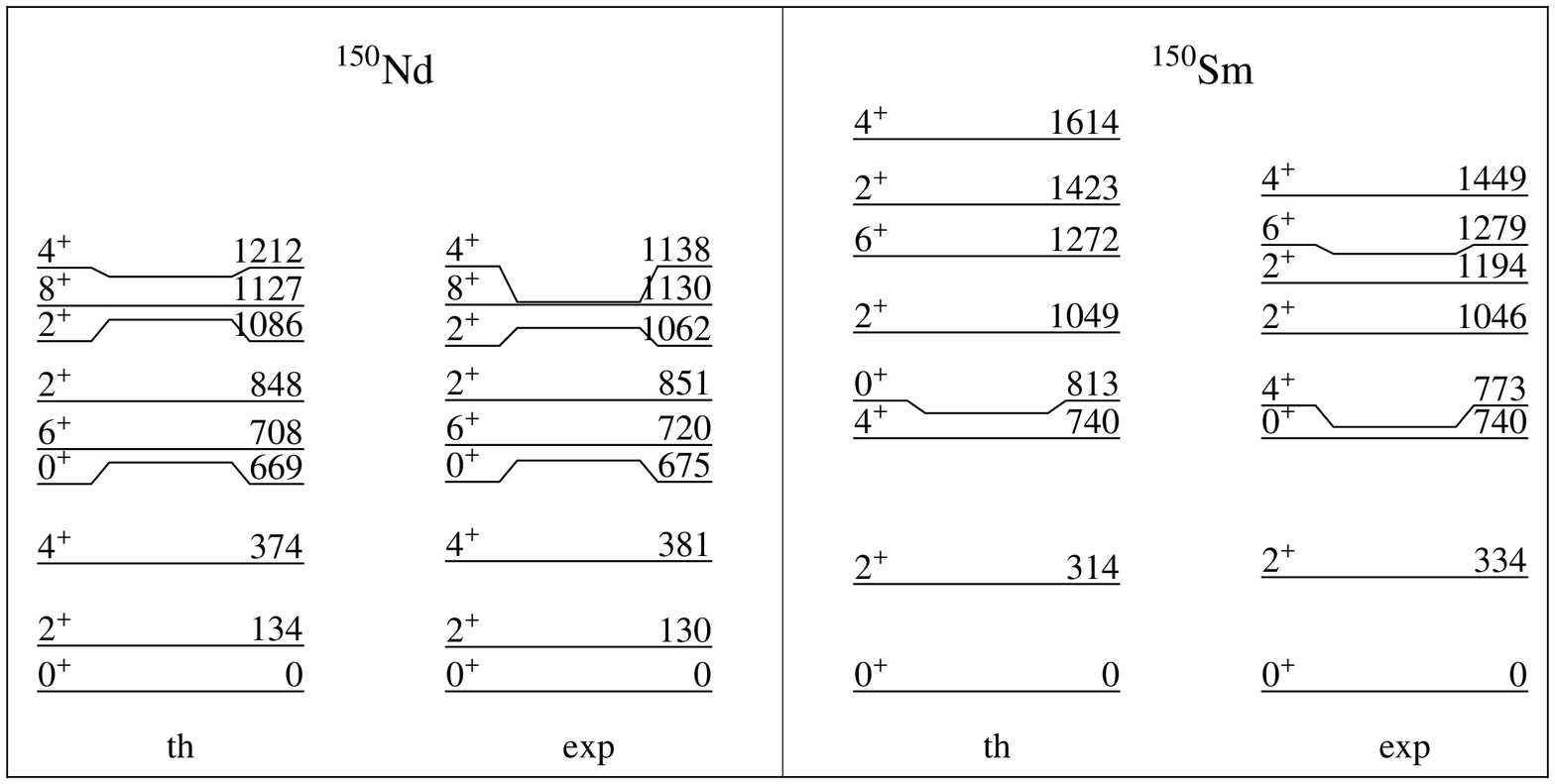} 
\end{center}
\caption{\label{spectra}Comparison between calculated and experimental low-lying spectra for the pair of nuclei $^{150}$Nd and $^{150}$Sm.}
\end{figure*}

In order to evaluate the matrix elements we make use of the microscopic
interacting boson model (IBM-2) \cite{iac1}. The method of evaluation
is discussed in detail in \cite{barea}. Here we briefly mention the
logic of the method, which  is a mapping of the fermion operator $H$
onto a boson space and its evaluation with bosonic wave functions.
The mapping \cite{otsuka} can be done to leading order (LO), next
to leading order (NLO), etc. In Ref.~\cite{barea} we showed, by
explicit calculations, that NLO terms give, in general, negligible
contribution, $\leq1\%$. In this article, we present only LO calculations.
The matrix elements of the mapped operators are then evaluated with
realistic wave functions, taken either from the literature, when available,
or obtained from a fit to the observed energies and other properties
($B(E2)$ values, quadrupole moments, $B(M1)$ values, magnetic moments, etc.).
The values of the parameters used in the calculation are given in
the appendices. In Appendix A, we give the neutrino potential and
its parameters. In Appendix B, we list the single-particle and -hole
energies and strengths of interaction. In Appendix C, we give the parameters
of the IBM-2 Hamiltonian for each nucleus considered in this article,
together with their references. As shown in the references quoted in Appendix C, the quality of the IBM-2 wave functions ranges from very good to excellent for nuclei with $A \gtrsim 70$ where collective features are very pronounced, especially in deformed nuclei. As an example, we show in Fig.~\ref{spectra} a comparison between calculated and experimental spectra for the pair of nuclei $^{150}$Nd and $^{150}$Sm.
For nuclei with $A \lesssim 70$, the IBM-2 description is only approximate, and one needs to go to the isospin conserving versions IBM-3 \cite{ell80} and IBM-4 \cite{ell81}. Nonetheless, we will report, for the sake of completeness, also results for $^{48}$Ca decay, with the proviso that these are rather approximate. Also, in some cases, intruder configurations play a role, especially in the structure of the excited $0^+$ state, and one needs to go to the configuration-mixing version IBM2-CM \cite{duv82}. All these improvements will be reported in subsequent papers.

\subsection{Results}

The matrix elements of the operator $H(p)$ have dimension fm$^{-1}$.
It has become customary to quote the values of $M^{(0\nu)}$ by multiplying
by the nuclear radius in fm, $R=R_{0}A^{1/3}$, with $R_{0}=1.2$~fm.
The matrix elements are then dimensionless.

\subsubsection{0$\nu\beta\beta$ decay with light neutrino exchange}
\begin{table*}[cbt!]
 \caption{\label{table3}IBM-2 nuclear matrix elements $M^{(0\nu)}$ (dimensionless) for $0\nu\beta\beta$ decay with Jastrow M-S SRC and $g_V/g_A=1/1.269$.}
 \begin{ruledtabular} %
\begin{tabular}{ccccccccc}
&\multicolumn{4}{c}{$0_{1}^{+}$} &\multicolumn{4}{c}{$0_{2}^{+}$}\\ \cline{2-5} \cline{6-9}
\T
$A$  & $M_{GT}^{(0\nu)}$  & $M_{F}^{(0\nu)}$  & $M_{T}^{(0\nu)}$  & $M^{(0\nu)}[0_{1}^{+}]$  & $M_{GT}^{(0\nu)}$  & $M_{F}^{(0\nu)}$  & $M_{T}^{(0\nu)}$  & $M^{(0\nu)}[0_{2}^{+}]$\tabularnewline
\hline 
\T
76  & 4.10  & -2.53  & -0.25  & 5.42  & 1.81  &-1.21   & -0.10  & 2.46\tabularnewline
82  & 3.26  & -2.12  & -0.25  & 4.37 & 0.86  &-0.69   & -0.05  & 1.23\tabularnewline
\hline
96   & 2.26  & -0.24  & 0.13  & 2.53  & 0.04  &-0.00   & 0.00  & 0.04\tabularnewline
100  & 3.32  & -0.33  & 0.20  & 3.73  & 0.88  &-0.09   & 0.05  & 0.99\tabularnewline
110  & 3.22  & -0.26  &0.24   & 3.62  & 0.41  &-0.04   & 0.03  & 0.46\tabularnewline
116  & 2.49  & -0.23   &0.15   & 2.78 & 0.78  &-0.06   & 0.04  & 0.85\tabularnewline
\hline
124  & 2.69  & -1.53  & -0.13  & 3.50  & 2.03 & -1.23  & -0.10  & 2.70\tabularnewline
128  & 3.46  & -1.90  & -0.16  & 4.48  & 2.44 & -1.40  & -0.10  & 3.22\tabularnewline
130  & 3.12  & -1.69  & -0.14  & 4.03  & 2.33 & -1.32  & -0.09  & 3.07\tabularnewline
136  & 2.59  & -1.37  & -0.11  & 3.33  & 1.40 &-0.75   & -0.04  & 1.82\tabularnewline
\hline
148  & 1.73  & -0.28  & 0.08  & 1.98  & 0.22  & -0.04  & 0.01  & 0.25\tabularnewline
150  & 2.03  & -0.28  & 0.11  & 2.32  & 0.35  &-0.05   & 0.02  & 0.39\tabularnewline
154  & 2.23  & -0.26  & 0.12  & 2.50  & 0.01  &-0.01   & 0.01  & 0.02\tabularnewline
160  & 3.25  & -0.31  & 0.18  & 3.62  & 0.66  &-0.08   & 0.05  & 0.75\tabularnewline
198  & 1.64  & -0.23  & 0.10  & 1.88  & 0.07  &-0.01   & 0.01  & 0.08\tabularnewline
\hline
48  & 1.53  & -1.03  & -0.19  & 1.98   & 3.62  & -3.78  & -0.13  & 5.83\tabularnewline
\end{tabular}\end{ruledtabular} 
\end{table*}

\begin{ruledtabular}
\begin{center}
\begin{table}[cbt!]
\caption{\label{table4}
Nuclear matrix elements for $0\nu\beta\beta$ decay to the ground state, $0_{1}^{+}$, in IBM-2 with Jastrow M-S SRC and $g_A=1.269$, QRPA-T\"u with Jastrow M-S SRC and $g_A=1.254$ \cite{simkovic1}, and the ISM with Jastrow M-S SRC and $g_A=1.25$ \cite{poves}. All matrix elements are in dimensionless units.}
\begin{tabular}{cccc}
Decay   &   &  \ensuremath{M^{(0\nu)}} &  \\ \cline{2-4}
\T
 &  IBM-2  &  QRPA-T\"u  &  ISM \\ 
\hline
\T
$^{48}$Ca$\rightarrow ^{48}$Ti		&1.98	&	 	&0.54\\
$^{76}$Ge$\rightarrow ^{76}$Se		&5.42	&4.68 	&2.22\\
$^{82}$Se$\rightarrow ^{82}$Kr		&4.37	&4.17 	&2.11\\
$^{96}$Zr$\rightarrow ^{96}$Mo		&2.53	&1.34 	&\\
$^{100}$Mo$\rightarrow ^{100}$Ru 	&3.73	&3.53 	&\\
$^{110}$Pd$\rightarrow ^{110}$Cd 	&3.62	&	 	&\\
$^{116}$Cd$\rightarrow ^{116}$Sn 	&2.78	&2.93 	&\\
$^{124}$Sn$\rightarrow ^{124}$Te 	&3.50	&	 	&2.02\\
$^{128}$Te$\rightarrow ^{128}$Xe 	&4.48	&3.77	&2.26\\
$^{130}$Te$\rightarrow ^{130}$Xe 	&4.03	&3.38	&2.04\\
$^{136}$Xe$\rightarrow ^{136}$Ba 	&3.33	&2.22	&1.70\\
$^{148}$Nd$\rightarrow ^{148}$Sm 	&1.98	&		&\\
$^{150}$Nd$\rightarrow ^{150}$Sm 	&2.32	&		&\\
$^{154}$Sm$\rightarrow ^{154}$Gd 	&2.50	&		&\\
$^{160}$Gd$\rightarrow ^{160}$Dy 	&3.62	&		&\\
$^{198}$Pt$\rightarrow ^{198}$Hg 	&1.88	&		&\\
\end{tabular}
\end{table}
\end{center}
\end{ruledtabular}

In Table~\ref{table3}, we show the results of our calculation of the matrix
elements to the ground state, $0_{1}^{+}$, and first excited state,
$0_{2}^{+}$, broken down into GT, F, and T contributions and their
sum according to Eq.~(19). We note that since we are covering all nuclei from $A=76$ to $A=198$, we have two classes of nuclei, those in which protons and neutrons occupy the same major shell ($A=76, 82, 124, 128, 130, and 136$) and those in which they occupy different major shells ($A=96, 100, 110, 116, 148, 150, 154, 160, and 198$). For example in $^{76}_{32}$Ge$_{44} \rightarrow^{76}_{34}$Se$_{42}$ decay both protons and neutrons occupy the shell 28-50, while in $^{110}_{46}$Pd$_{64} \rightarrow^{110}_{48}$Cd$_{62}$ decay protons occupy the shell 28-50, and neutrons occupy the shell 50-82. The magnitude of the Fermi matrix element which is related to the overlap of the proton and neutron wave functions is therefore different in these two classes of nuclei, being large in the former and small in the latter case. This implies a considerable amount of isospin violation for nuclei in the first class. The two classes are separated by lines in Table~\ref{table3} in order to make the distinction clear. For completeness, we have added at the bottom of the table the IBM-2 calculation of $^{48}$Ca$\rightarrow ^{48}$Ti, assuming $^{48}$Ca to be double magic. IBM-2 is rather poor in this case, as evidenced by the large Fermi matrix element, and the values in the table for $^{48}$Ca$\rightarrow ^{48}$Ti decay should be considered a rough estimate. Table~\ref{table3} also shows the tensor matrix elements, $M_T^{(0\nu)}$. These are systematically small (about 5\% of $M_{GT}^{(0\nu)}$) and have sign opposite to or the same as $M_{GT}^{(0\nu)}$ when protons and neutrons occupy the same major shell or not, respectively. This behavior can be traced to the fact that the neutrino potential $V(r)$ is different for the tensor contribution than for Fermi and Gamow-Teller contributions. In the notation of Table VIII of Ref.~\cite{barea}, $V(r)=H(r)$ for Fermi and Gamow-Teller matrix elements and $V(r)=-rH'(r)$ for tensor matrix elements.

A point of great interest is the comparison among various model calculations of the NME. Up to 2009, the methods used were the QRPA and the ISM. In addition to these, there are now our calculation (IBM-2) and calculations based on the density functional theory (DFT). Among the QRPA calculations there are those of the T\"ubingen group and those of the Jyv\"askyl\"a group. These calculations often use different parametrizations of the SRC, may or may not include $g_A^2$ in $M^{(0\nu)}$, use different values of $g_A$, and are done in the closure or non-closure approximation. Therefore, the comparison among matrix elements in different models should be taken only as relative to a given matrix element, for example, $^{76}$Ge $\rightarrow ^{76}$Se.

\begin{figure*}[cbt!]
\begin{center}
\includegraphics[width=12.cm]{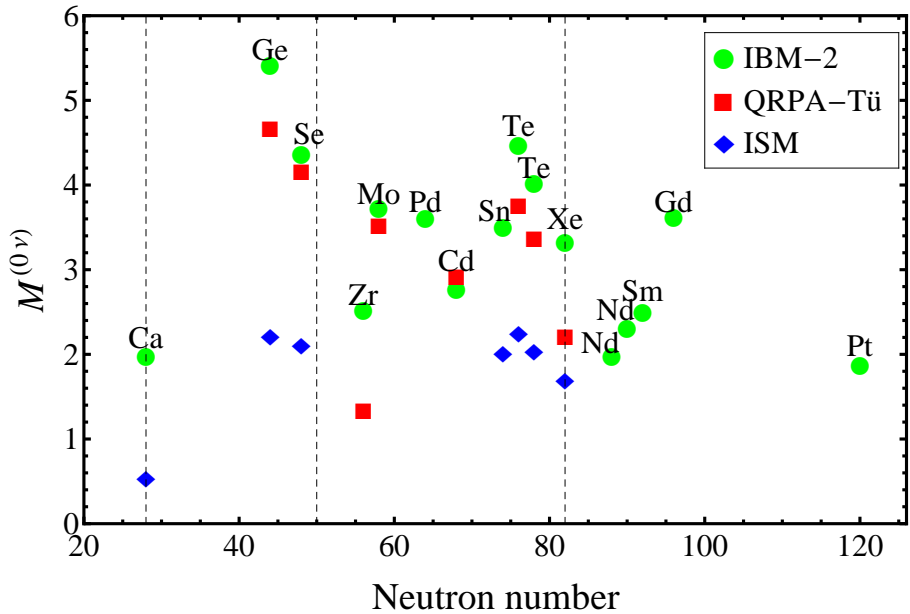} 
\end{center}
\caption{\label{fig4}(Color online) IBM-2 results for $0\nu\beta\beta$ compared with QRPA-T\"u \cite{simkovic1} and the ISM \cite{poves}.}
\end{figure*}

In Table~\ref{table4} we compare our results with those of a particular QRPA calculation, QRPA-T\"u \cite{simkovic1} and of an ISM calculation \cite{poves} with Miller-Spencer (M-S) parametrization of the SRC. The IBM-2 and QRPA-T\"u results show a similar variation with $A$, while the ISM results are, apart from the small value in $^{48}$Ca, clustered around $\sim 2.00$ in the entire range $A=76-136$, and are a factor of approximately 2 smaller than results from IBM-2 and QRPA-T\"u. It should be noted that, due to the different approximation made in each model, a range of values would be more appropriate. For example, if we set to zero the Fermi matrix element in our calculated $^{48}$Ca $\rightarrow ^{48}$Ti decay, we obtain $M^{(0\nu)}=1.33$ and thus our matrix elements should be more appropriately quoted as $M^{(0\nu)}=1.33-1.98$. The sensitivity of the IBM-2 results to parameter changes, model assumptions and operator assumptions is further discussed in Sec.~\ref{sensitivity}.
The results in Table~\ref{table4} are summarized in Fig.~\ref{fig4}, where they are plotted as a function of neutron number. The reason for this way of plotting is due to shell effects, as discussed in Sec. VI B of Ref.~\cite{barea}.
 The matrix elements $M^{(0\nu)}$ attain their smallest values at the closed proton and neutron shells due to the form of the transition operator which for $\beta^-\beta^-$ decay annihilates a neutron pair  and creates a proton pair.
 These shell effects are very clear in both the IBM-2 and the QRPA calculations, and to some extent also in the ISM calculation. They are responsible for the small matrix element in the decay of the doubly magic nucleus $^{48}$Ca. They are also responsible for the ratio of the matrix elements of two different isotopes of the same element. For example, a simple calculation using the pair operators of Eq.~(42) of Ref.~\cite{barea},  gives $M^{(0\nu)}(^{128}$Te$)/M^{(0\nu)}(^{130}$Te$)=1.11$, to be compared with 1.11 from IBM-2,   1.13 from QRPA-T\"u, and 1.11 from ISM.

\begin{ruledtabular}
\begin{center}
\begin{table}[cbt!]
\caption{\label{table5}Neutrinoless nuclear matrix elements to the first excited state, $0_{2}^{+}$, in IBM-2, QRPA-T\"u RCM/BEM \cite{simkovic2001}, and the ISM \cite{menendez2009}. All matrix elements are in dimensionless units.}
\begin{tabular}{cccc}
Decay   &   &  \ensuremath{M^{(0\nu)}} &  \\ \cline{2-4}
\T
 &  IBM-2  &  QRPA-T\"u  &  ISM \\ 
\hline
 \T
$^{48}$Ca$\rightarrow ^{48}$Ti		&5.83	&	 	&0.68\\
$^{76}$Ge$\rightarrow ^{76}$Se		&2.46	&1.28/0.99 	&1.49\\
$^{82}$Se$\rightarrow ^{82}$Kr	&1.23$^*$	&1.34/0.95$^*$ 	&0.28$^*$\\
$^{96}$Zr$\rightarrow ^{96}$Mo		&0.04	&	 	&\\
$^{100}$Mo$\rightarrow ^{100}$Ru 	&0.99	&1.27/1.76 	&\\
$^{110}$Pd$\rightarrow ^{110}$Cd 	&0.46	&	 	&\\
$^{116}$Cd$\rightarrow ^{116}$Sn 	&0.85	&	 	&\\
$^{124}$Sn$\rightarrow ^{124}$Te 	&2.70	&	 	&0.80\\
$^{128}$Te$\rightarrow ^{128}$Xe 	&3.22$^*$	&		&\\
$^{130}$Te$\rightarrow ^{130}$Xe 	&3.07	&		&0.19\\
$^{136}$Xe$\rightarrow ^{136}$Ba 	&1.82	&4.42/0.44	&0.49\\
$^{148}$Nd$\rightarrow ^{148}$Sm 	&0.25	&		&\\
$^{150}$Nd$\rightarrow ^{150}$Sm 	&0.39	&		&\\
$^{154}$Sm$\rightarrow ^{154}$Gd 	&0.02	&		&\\
$^{160}$Gd$\rightarrow ^{160}$Dy 	&0.75	&		&\\
$^{198}$Pt$\rightarrow ^{198}$Hg 	&0.08$^*$	&		&\\
\end{tabular}
* Negative $Q$-value
\end{table}
\end{center}
\end{ruledtabular}

\begin{figure}[cbt!]
\begin{center}
\includegraphics[width=8.6cm]{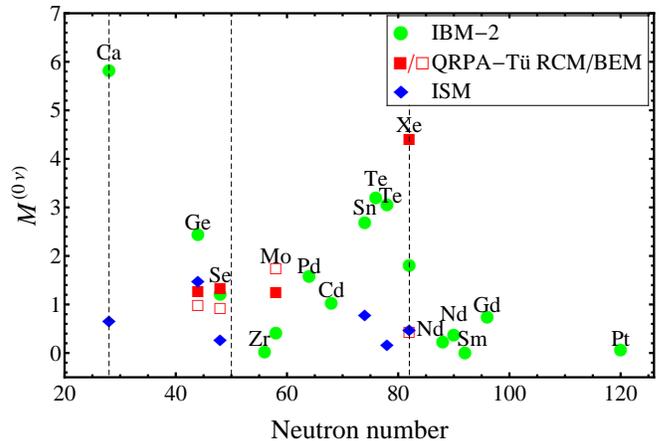} 
\end{center}
\caption{\label{fig5}(Color online) IBM-2 results for $0\nu\beta\beta$ decay to $0_{2}^{+}$ compared with QRPA-T\"u \cite{simkovic2001} and the ISM \cite{menendez2009}.}
\end{figure}

Our results to $0^+_2$ are shown in Table \ref{table5}. Because of the reduced phase-space factor for decay to $0_{2}^{+}$,
this table is of less interest. In this case, there appears to be no
correlation between IBM-2 and other calculations. It should be noted, however, that the QRPA-T\"u results shown in Table~\ref{table5} were done before an error was discovered in the treatment of short-range correlations \cite{!!} and with two different methods for treating the excited $0^+_2$ state, the recoupling method (RCM) and the boson expansion method (BEM) \cite{simkovic2001}. These results are therefore inconsistent by a factor of approximately 2 with those in Table~\ref{table4} and Ref.~\cite{simkovic1} for $0^+_1$. Also, IBM-2 calculations
have been done without including intruder configurations. It is known
that, in some cases, the $0_{2}^{+}$ state is an intruder state.
The most notable cases are Ge, Mo, Cd, Nd, and Hg isotopes \cite[p.180]{iac11}.
Although configuration-mixing IBM-2 calculations for these nuclei
are available, they have not been implemented yet in the calculation
of 0$\nu\beta\beta$ to $0_{2}^{+}$ states. The comparison among IBM-2, QRPA-T\"u, and the ISM is shown in Fig.~\ref{fig5}.

The most detailed comparison among different model calculations yet has been done recently by Suhonen \cite{suh}. This author has shown a very close correspondence between the IBM-2 results and the QRPA-Jy result, Table~\ref{newtabla}, and has argued that the reason why the QRPA and IBM-2 agree is because the QRPA can be seen as a leading-order boson expansion. This statement should be taken, however, with caution since the QRPA results require the adjustment of the parameter $g_{pp}$.
\begin{ruledtabular}
\begin{center}
\begin{table}[h]
 \caption{\label{newtabla}Comparison between IBM-2 and QRPA-Jy \cite{suh} nuclear matrix elements $M^{(0\nu)}$ (dimensionless) for $0\nu\beta\beta$ decay. 
 }
\begin{tabular}{ccccc}
A &\multicolumn{2}{c}{$0_{1}^{+}$} &\multicolumn{2}{c}{$0_{2}^{+}$}\\ \cline{2-3} \cline{4-5}
\T
  & IBM-2 &QRPA & IBM-2 &QRPA\tabularnewline
\hline 
48  & 1.98  	&1.09-1.89  	& 5.83	&  	\tabularnewline
76  & 5.42  	&2.28-4.17  	& 2.46	&2.47-5.38  	\tabularnewline
82  & 4.37 	&2.11-3.51   	& 1.23	&0.831-1.85  	\tabularnewline
96   & 2.53  &2.00-2.07  	& 0.04	&1.96  	\tabularnewline
100  & 3.73  &2.26-2.74   	& 0.99	&0.31  	\tabularnewline
110  & 3.62  &3.63-4.51   	& 0.46	&0.96-1.73  	\tabularnewline
116  & 2.78 	&2.36-3.98  	& 0.85	&0.25  	\tabularnewline
124  & 3.50  &2.58-4.18   	& 2.70	&3.96-5.88  	\tabularnewline
128  & 4.48  &2.74-4.15   	& 3.22	&  	\tabularnewline
130  & 4.03  &2.60-3.78   	& 3.07	&3.88-6.61  	\tabularnewline
136  & 3.33  &1.83-2.53   	& 1.82	&2.75-6.08  	\tabularnewline
\end{tabular}
\end{table}
\end{center}
\end{ruledtabular}

Another question which has been extensively analyzed in recent months is the size of the Fermi matrix elements, and its comparison among different models. To this end, it is convenient to introduce the quantity $\chi_F=(g_V/g_A)^2 M^{(0\nu)}_F/M^{(0\nu)}_{GT}$. This quantity is shown in Table~\ref{chitable}. One can see that the situation is more complex than in the case of the overall matrix elements, due to the different approximations made by the different models. The IBM-2 results are large for nuclei in which protons and neutrons occupy the same major shells, but small in cases in which protons and neutrons occupy different major shells. For QRPA-Jy they are uniformly large ($\chi_F \sim -0.30$) while for the ISM they are uniformly small ($\chi_F \sim -0.15$).

The large Fermi matrix elements in IBM-2 with protons and neutrons in the same major shell and in the QRPA throughout point to large isospin violation in the wave functions of the initial and final nuclei. In the case of $2\nu\beta\beta$ decay, if isospin is a good quantum number, the Fermi matrix elements should identically vanish. By a similar argument, the Fermi matrix elements in $0\nu\beta\beta$ are expected to be small, although not zero, the main difference between $2\nu\beta\beta$ and $0\nu\beta\beta$ being the neutrino potential, given in Appendix A, Table~\ref{table16}. For this reason, the calculated values of $\chi_F$ given in Table~\ref{chitable} may be entirely spurious and should be considered with a large error. It is difficult to estimate the error in $\chi_F$ introduced by isospin violation, since there are no direct experimental data for single-$\beta$ $0^+ \rightarrow 0^+$ transitions from odd-odd to even-even (or vice versa) nuclei in heavy nuclei. The estimate of the error also depends on which nucleus is considered. The maximum error is 100\%, if the Fermi matrix elements are entirely spurious, in which case the values of Table~\ref{chitable} should be quoted as $-0.42(42)$ for $^{48}$Ca and similarly for all others. Another estimate is to extract the error by comparing with the ISM calculations which have the smallest values of $\chi_F$, in which case the quoted value should be $-0.42(28)$ with an error of 67\%. We have used this estimate of the error in the following subsection 3 and in Table~\ref{final}.
\begin{ruledtabular}
\begin{center}
\begin{table}[h]
\caption{\label{chitable}Comparison among Fermi matrix elements, $\chi_F$, in IBM-2, QRPA-Jy \cite{suhpriv} and the ISM \cite{caurier2007}.}
\begin{tabular}{cccc}
Decay &\multicolumn{3}{c}{$\chi_F$} \\ \cline{2-4}
\T
      		&  IBM-2 &  QRPA-Jy   &  ISM \\
 \hline
\T
$^{48}$Ca		&-0.42		&-0.56\footnotemark[1]		&-0.14	\\
$^{76}$Ge		&-0.38		&-0.22		&-0.10	\\
$^{82}$Se		&-0.42		&-0.28		&-0.10	\\
$^{96}$Zr		&-0.06		&-0.43		&	\\
$^{100}$Mo		&-0.06		&-0.40		&	\\
$^{110}$Pd		&-0.05		&-0.38		&-0.16	\\
$^{116}$Cd		&-0.06		&-0.28		&-0.19	\\
$^{124}$Sn		&-0.35		&-0.42		&-0.13	\\
$^{128}$Te		&-0.34		&-0.37		&-0.13	\\
$^{130}$Te		&-0.34		&-0.37		&-0.13	\\
$^{136}$Xe		&-0.33		&-0.34		&-0.13	\\
$^{148}$Nd		&-0.10		&			&	\\
$^{150}$Nd		&-0.09		&			&	\\
$^{154}$Sm		&-0.07		&			&	\\
$^{160}$Gd		&-0.06		&			&	\\
$^{198}$Pt		&-0.09		&			&	\\
\end{tabular}
\footnotetext[1]{Reference~\cite{suh93}.}
\end{table}
\end{center}
\end{ruledtabular}

In addition to the calculations discussed above, several others have been made, most notably in the deformed QRPA \cite{fan10,fan11} and in the projected Hartree-Fock-Bogoliubov (HFB) framework \cite{chandra}, and using the energy density functional method \cite{Martinez-Pinedo}. Since these use SRC with Argonne/CD-Bonn and the unitary correlation method (UCOM) they will be discussed at the end of Sec.~\ref{sensitivity}.

\subsubsection{$0\nu\beta\beta$ decay with heavy neutrino exchange}

\begin{table*}[cbt!]
 \caption{\label{table6}Nuclear matrix elements for the heavy neutrino exchange mode of the
neutrinoless double-$\beta$ decay to the ground state (columns 2, 3, 4,
and 5) and to the first excited state (columns 6, 7, 8, and 9) using
the microscopic interacting boson model (IBM-2) with Jastrow M-S SRC and $g_V/g_A=1/1.269$.}
 \begin{ruledtabular} %
\begin{tabular}{ccccccccc}
$A$ &\multicolumn{4}{c}{$0_{1}^{+}$} &\multicolumn{4}{c}{$0_{2}^{+}$}\\ \cline{2-5} \cline{6-9}
\T
  & $M_{GT_h}^{(0\nu)}$  & $M_{F_h}^{(0\nu)}$  & $M_{T_h}^{(0\nu)}$  & $M_{h}^{(0\nu)}[0_{1}^{+}]$  & $M_{GT_h}^{(0\nu)}$  & $M_{F_h}^{(0\nu)}$  & $M_{T_h}^{(0\nu)}$  & $M_{h}^{(0\nu)}[0_{2}^{+}]$\tabularnewline
\hline 
\T
76  & 52.5  &-39.5   & -29.0  & 48.1  & 19.5  &-15.4   & -10.5  & 18.5\tabularnewline
82  & 43.9  &-33.9   & -29.4  & 35.6  & 8.53  &-7.36   & -4.96  & 8.14\tabularnewline
\hline
96  & 33.6   &-18.8  & 13.6  & 59.0  & 0.700  &-0.382   & 0.321  & 1.26\tabularnewline
100  & 54.5   &-30.0  & 26.1  & 99.3  & 4.59  &-2.55   & 1.70  & 7.87\tabularnewline
110  & 49.3  &-26.5   & 29.9  & 95.7  & 18.4  &-9.73   & 10.0  & 34.5\tabularnewline
116  & 36.4  &-20.1   & 18.2  & 67.1  & 13.9  &-7.72   & 7.07  & 25.8\tabularnewline
\hline
124  & 37.0  &-27.1   & -16.1  & 37.8  & 25.2  &-19.0   & -10.6  & 26.4\tabularnewline
128  & 47.0  &-34.1   & -19.7  & 48.4  & 28.0  &-21.1   & -11.1  & 30.0\tabularnewline
130  & 42.7  &-30.9   & -17.9  & 44.0  & 25.9  &-19.6   & -10.0  & 28.1\tabularnewline
136  & 33.7  &-24.3   & -13.7  & 35.1  & 12.9  &-9.90   & -4.39  & 14.7\tabularnewline
\hline
148  & 35.8  &-21.2   & 10.4  & 59.4  & 3.99  &-2.43   & 1.07  & 6.57\tabularnewline
150  & 39.9  &-23.0   & 14.2  & 68.4  & 5.90  &-3.42   & 1.90  & 9.92\tabularnewline
154  & 38.6  &-21.8   & 14.9  & 67.1  & 1.93  &-1.12   & 1.20  & 3.82\tabularnewline
160  & 51.9  &-28.7   & 23.2  & 92.9  & 13.1  &-7.34   & 6.65  & 24.3\tabularnewline
198  & 34.7  &-20.8   & 13.8  & 61.5  & 1.51  &-0.899   & 0.688  & 2.76\tabularnewline
\hline
48  & 26.8  &-20.1   & -22.9  & 16.3  & 23.8  &-27.3   & -6.95  & 33.8\tabularnewline
\end{tabular}\end{ruledtabular} 
\end{table*}
\begin{ruledtabular}
\begin{center}
\begin{table}[cbt!]
\caption{\label{table7}Neutrinoless nuclear matrix elements to the ground state, $0_{1}^{+}$, in IBM-2 and QRPA-T\"u \cite{simkovic} for the heavy neutrino exchange mode and M-S SRC. All matrix elements are in dimensionless units.}
\begin{tabular}{ccc}
Decay &\multicolumn{2}{c}{  \ensuremath{M_{h}^{(0\nu)}}}   \\ \cline{2-3}
\T
  &  IBM-2  &  QRPA-T\"u   \\
 \hline
 \T
$^{48}$Ca$\rightarrow ^{48}$Ti		&16.3	&	 	\\
$^{76}$Ge$\rightarrow ^{76}$Se		&48.1	&32.6 	\\
$^{82}$Se$\rightarrow ^{82}$Kr		&35.6	&30.0 	\\
$^{96}$Zr$\rightarrow ^{96}$Mo		&59.0	&14.7	 	\\
$^{100}$Mo$\rightarrow ^{100}$Ru 	&99.3	&29.7 	\\
$^{110}$Pd$\rightarrow ^{110}$Cd 	&95.7	&	 	\\
$^{116}$Cd$\rightarrow ^{116}$Sn 	&67.1	&21.5	 	\\
$^{124}$Sn$\rightarrow ^{124}$Te 	&37.8	&	 	\\
$^{128}$Te$\rightarrow ^{128}$Xe 	&48.4	&26.6		\\
$^{130}$Te$\rightarrow ^{130}$Xe 	&44.0	&23.1		\\
$^{136}$Xe$\rightarrow ^{136}$Ba 	&35.1	&14.1	\\
$^{148}$Nd$\rightarrow ^{148}$Sm 	&59.4	&		\\
$^{150}$Nd$\rightarrow ^{150}$Sm 	&68.4	&35.6		\\
$^{154}$Sm$\rightarrow ^{154}$Gd 	&67.1	&		\\
$^{160}$Gd$\rightarrow ^{160}$Dy 	&92.9	&		\\
$^{198}$Pt$\rightarrow ^{198}$Hg 	&61.5	&		\\
\end{tabular}
\end{table}
\end{center}
\end{ruledtabular}
These matrix elements can be simply calculated by replacing the potential
$v(p)=2\pi^{-1}[p(p+\tilde{A})]^{-1}$ of Eq.~(\ref{eq:pot_light})
with the potential $v_{h}(p)=2\pi^{-1}(m_{e}m_{p})^{-1}$ of
Eq.~(\ref{eq:pot_heavy}). Table~\ref{table6} gives the corresponding matrix
elements. The index $h$ distinguishes
these matrix elements from those with light neutrino exchange. 
Our results are compared with QRPA-T\"u results in Table~\ref{table7}. These QRPA results
are obtained with Jastrow SRC \cite{simkovic} and prior to the more refined treatment of SRC of Ref.~\cite{simkovic1}, and are shown here for the sake of comparison of the $A$ dependence, not of their absolute magnitude, which is a factor of approximately 2 smaller than in IBM-2.
It has been suggested that measurement of neutrinoless double-$\beta$
decay in different nuclei may be used to distinguish between the two
mechanisms, light or heavy neutrino exchange. However, the results
in Table~\ref{table3} and Table~\ref{table7} are highly correlated as is clear from
the fact that they are obtained one from the other by replacing the
potential $v(p)$ with $v_{h}(p)$. Therefore, this criterion cannot
be used to distinguish between the two mechanisms \cite{lisi}.

\subsubsection{Sensitivity to parameter changes, model assumptions and operator
assumptions}
\label{sensitivity}
Many ingredients go into the calculation of the nuclear matrix elements.
In Ref.~\cite{barea}, the sensitivity to input parameter changes
was estimated from the sensitivity to parameter changes in five quantities: (1) single-particle energies; (2) strengths of interactions; (3)
oscillator parameter in the single-particle wave functions; (4) closure energy in the neutrino potential; and (5) nuclear radius. (1) The sensitivity to single-particle energies
was emphasized in Ref.~\cite{suhonen} within the framework of the QRPA
and has been the subject of several experimental investigations \cite{schiffer,kay}.
We estimate it to be 10\%. (2) We estimate the sensitivity strengths of interactions, in the present
case the surface delta interaction used to calculate the structure
of the pair states. We estimate this to be 5\%. We note that this
is the main source of sensitivity in the QRPA, especially in nuclei close
to the spherical-deformed transition, for example $^{150}$Nd. (3) We estimate the sensitivity to the
oscillator parameer to be 5\%. (4)  We estimate the sensitivity to closure energy to be 5\%. (5) If the matrix elements are quoted
in dimensionless units there is also the sensitivity to $R$. We estimate
this to be 5\%. However, this sensitivity can be reduced to a small
value, 1\%, if the experimental rms value is used instead of the formula
$R=R_{0}A^{1/3}$. The total estimated sensitivity to input parameters
is 30\% if all contributions are added or 14\% if combined in quadrature.

In addition, we estimate the sensitivity to model assumptions to be:
(1) truncation to pairs with angular momentum $J=0$ and $J=2$ (S-D space) and (2) isospin purity. We estimate the former to range from 1\%
in spherical nuclei to 10\% in deformed nuclei.
For the latter we estimate this to be small, 1\%, for GT and T matrix elements, and
large, 10\%, for F matrix elements. Taking into account the fact that
the F matrix elements contribute only $\sim$20\% to the total matrix
elements, we estimate the total sensitivity to model assumptions to
be ranging from 3\% in spherical nuclei, to 12\% in deformed nuclei
(in addition) or 2\%-10\% (in quadrature). A special case is that of $^{48}$Ca decay for which, for reasons mentioned above, the Fermi matrix elements are overestimated. In this case the sensitivity to model assumptions may be as high as 20\% (in addition) or 16\% (in quadrature).

\begin{table*}[cbt!]
 \caption{\label{table9}Comparison among matrix elements $M^{(0\nu)}$ calculated with Miller-Spencer (M-S) and Argonne/CD-Bonn (CCM)
in IBM-2, with M-S, CCM, and UCOM in QRPA-T\"u \cite{simkovic1, argonnebonn,fan11}, and with M-S and UCOM in the ISM \cite{poves, menendez2009}. Note that the QRPA matrix elements are evaluated using $g_A=1.254$ and the ISM matrix elements are evaluated using $g_A=1.25$.}
 \begin{ruledtabular} 
\begin{tabular}{cccccccc}
A	&\multicolumn{2}{c}{IBM-2}     &\multicolumn{3}{c}{QRPA-T\"u}    &\multicolumn{2}{c}{ISM}  \\ \cline{2-3}
 \cline{4-6} \cline{7-8} 
\T
 	&  M-S  &  CCM  &  M-S\footnotemark[1]  &  CCM\footnotemark[2] & UCOM\footnotemark[1]  &  M-S\footnotemark[3]  &  UCOM\footnotemark[4] \\
\hline 
\T
48  & 1.98  & 2.28/2.38  &		&	&				&0.59		&0.85\tabularnewline
76  & 5.42  & 5.98/6.16  &4.68	&5.81/6.32	&5.73	&2.22		&2.81\tabularnewline
82  & 4.37  & 4.84/4.99  &4.17	&5.19/5.65	&5.09	&2.11		&2.64\tabularnewline
96  & 2.53  & 2.89/3.00  &1.34	&1.90/2.09	&1.79	&		&\tabularnewline
100  & 3.73  & 4.31/4.50 &3.52	&4.75/5.25	&4.58	&		& \tabularnewline
110  & 3.62  & 4.15/4.31 &		&	&				&		& \tabularnewline
116  & 2.78  & 3.16/3.29 &2.93	&3.54/3.99	&3.54	&		& \tabularnewline
124  & 3.50  & 3.89/4.02 &		&	&				&2.02	&2.62 \tabularnewline
128  & 4.48  & 4.97/5.13 &3.77	&4.93/5.49	&4.76	&2.26	&2.88 \tabularnewline
130  & 4.03  & 4.47/4.61 &3.38	&4.37/4.92	&4.26	&2.04	&2.65 \tabularnewline
136  & 3.33  & 3.67/3.79 &2.22	&2.78/3.11	&2.76	&1.70	&2.19 \tabularnewline
148  & 1.98  & 2.36/2.49 &		&	&				&		& \tabularnewline
150  & 2.32  & 2.74/2.88 &		&3.34\footnotemark[5]	&				&		& \tabularnewline
154  & 2.50  & 2.91/3.04 &		&	&				&		& \tabularnewline
160  & 3.62  & 4.17/4.34 &		&3.76\footnotemark[5]	&				&		& \tabularnewline
198  & 1.88  & 2.25/2.37 &		&	&				&		& \tabularnewline
\end{tabular}\end{ruledtabular} 
\footnotetext[1]{Reference~\cite{simkovic1}.}
\footnotetext[2]{Reference~\cite{argonnebonn}.}
\footnotetext[3]{Reference~\cite{poves}.}
\footnotetext[4]{Reference~\cite{menendez2009}.}
\footnotetext[5]{Reference~\cite{fan11}.}
\end{table*}

Finally, there is the estimated sensitivity to operator assumptions:
(1) Form of the transition operator. We have already commented in
Sec. II. A on the differences between \v{S}imkovic's \cite{simkovic} and
Tomoda's \cite{tomoda} formulations. This is a source of considerable
uncertainty. We estimate the sensitivity to 5\% by comparing our calculations
using Tomoda's and \v{S}imkovic's formulations. However, there still remains
the question of the recoil contribution which was a source of major
disagreement in early calculations. (2) Finite nuclear size (FNS).
We estimate this to be small, 1\%, due to the fact that we have used
realistic nucleon form factors with parameters determined from experiment.
(3) Short-range correlations (SRC). The sensitivity to the form of
short-range correlations has been very recently the subject of many
studies. To investigate this point, we have done calculations
with three different types of correlations, (a) Jastrow Miller-Spencer
(MS) and (b/c) Argonne/CD-Bonn (CCM). The results are shown in\
Table~\ref{table9}, 
 where we also show a comparison with results of calculations in the QRPA and the ISM using M--S, CCM and the unitary correlation operator method (UCOM) \cite{fel98}. It appears that in going from M--S 
to CCM or UCOM the matrix elements in all three methods (IBM-2, the QRPA, and the ISM) increase. In IBM-2 the multiplicative factor when going from M--S to CCM--Argonne is 1.10-1.20. In QRPA-T\"u it is 1.21-1.42 from M--S to CCM-Argonne and 1.21-1.33 from M--S to UCOM. In the ISM the factor is 1.25-1.30 from M--S to  UCOM. This multiplicative factor was taken into account in  Ref.~\cite{xx} when comparing IBM-2 matrix elements with the ISM. The discrepancy between IBM-2 and QRPA-T\"u multiplicative factors is not understood and should be investigated further. In Table~\ref{table9}, we have also added recent calculations in the deformed QRPA for decay of $^{150}$Nd and $^{160}$Gd \cite{fan11}. One may note that the correspondence between the QRPA and IBM-2 persists even in deformed nuclei. 

The total estimated sensitivity here
is 11\% (addition) or 7\% (quadrature), under the assumption of no recoil contribution
to the matrix elements.
Combining all contributions, we have a total estimated sensitivity of 44\%-55\%
if all the contributions are added or 16\%-19\% if they are combined
in quadrature.

The short-range correlations affect $0\nu_h\beta\beta$ decay differently than $0\nu\beta\beta$. We have therefore investigated the dependence of $0\nu_h\beta\beta$ matrix elements with M-S, and CCM correlations. The results are shown in Table~\ref{tablenew2}. We see here an increase of a factor from  1.69 to 2.80 when going from M--S to Argonne/CD-Bonn. This is because, as remarked in Appendix A, the neutrino potential for heavy neutrino exchange is a contact interaction in configuration space and thus strongly influenced by SRC. The correlation function in Eq. (\ref{jastrow}) has a value $f_J(0)=1-c$ at $r=0$. For the M--S parametrization $c=1$, $f_J(0)=0$, and thus, in the absence of a nucleon form factor, the matrix elements $M_h^{(0\nu)}$ vanish. The Argonne/CD-Bonn parametrizations have $c=0.92/0.46$ and thus a non-zero value at $r=0$. These results are modified by the presence of the nucleon form factors of Eq. (\ref{ffeq}), and the final results depend strongly on the choice of $g_V(p^2)$ and $g_A(p^2)$. From Table~\ref{tablenew2}, columns 4 and 5, it appears that the increase in the matrix elements when going from M--S to CCM-Argonne in QRPA-T\"u is much larger, from 7.01 to 10.1, than in IBM-2. As in the case of light neutrino exchange, this discrepancy is not understood and should be investigated further. The large increase both in IBM-2 and the QRPA also points to the strong sensitivity of the calculated $0\nu_h\beta\beta$ matrix elements to the specific form of SRC, and thus to the fact that the treatment here and in other calculations in the literature, through the nucleon form factors, may not be satisfactory. A more consistent treatment is discussed in Refs.~\cite{fae97, pre03}. In view of all these problems, we estimate the sensitivity to SRC for $0\nu_h\beta\beta$ decay to be much larger (50\%) than that for $0\nu\beta\beta$ decay (5\%). The total estimated sensitivity to operator assumptions for $0\nu_h\beta\beta$ is then 56\% (addition) and 50\% (quadrature).

\begin{table}[h]
 \caption{\label{tablenew2}Comparison among  $M^{(0\nu)}_h$ matrix elements calculated with different Jastrow
parametrizations for the SRC: Miller-Spencer and CCM
(Argonne and CD-Bonn) in IBM-2 and QRPA-T\"u \cite{simkovic, faessler2011}.}
 \begin{ruledtabular} %
\begin{tabular}{ccccc}
$A$  & \multicolumn{2}{c}{IBM-2}  & \multicolumn{2}{c}{QRPA-T\"u}\tabularnewline \cline{2-3} \cline{4-5}
\T
 & M-S  & CCM  & M-S\footnotemark[1] &CCM\footnotemark[2] \tabularnewline
\hline 
\T
48  &16.3		&46.3/76.0	&&		  \tabularnewline
76  &48.1		&107/163  	&32.6 &233/351 	  \tabularnewline
82  &35.6		&84.4/132 	&30.0 &226/340  	  \tabularnewline
96	&59.0		&99.0/135 	&14.7 &  	  \tabularnewline
100 &99.3		&165/224  	&29.7 &250/388 	  \tabularnewline
110 &95.7		&155/208  	&& 	  \tabularnewline
116 &67.1		&110/149  	&21.5 & 	  \tabularnewline
124 &37.8		&79.6/120 	&&  	  \tabularnewline
128 &48.4		&101/152  	&26.6 & 	  \tabularnewline
130 &44.0		&92.0/138 	&23.1 &234/364  	  \tabularnewline
136 &35.1		&72.8/109 	&14.1 &  	  \tabularnewline
148 &59.4		&103/142  	&& 	  \tabularnewline
150 &68.4		&116/160  	&35.6 & 	  \tabularnewline
154 &67.1		&113/155  	&& 	  \tabularnewline
160 &92.9		&155/211  	&& 	  \tabularnewline
198 &61.5		&104/141  	&& 	  \tabularnewline
\end{tabular}\end{ruledtabular} 
\footnotetext[1]{Reference~\cite{simkovic}.}
\footnotetext[2]{Reference~\cite{faessler2011}.}

\end{table}
To summarize the situation we show in Table~\ref{final} our final results with M-S SRC, together with an estimate of the error, based on the arguments given above. The error estimate is 30\% in $^{48}$Ca, 19\% in nuclei with protons and neutrons in the same major shell and 16\% in nuclei with protons and neutrons in different major shells, for $0\nu\beta\beta$. For $0\nu_h\beta\beta$, our estimated error is dominated by SRC. In Table~\ref{final} we have used 58\% in $^{48}$Ca, 53\% in nuclei with protons and neutrons in the same major shell and 52\% in nuclei with protons and neutrons in different major shells.

  \begin{table}[h]
 \caption{\label{final}Final IBM-2 matrix elements with M-S SRC and error estimate.}
 \begin{ruledtabular} %
\begin{tabular}{ccc}
Decay  & Light neutrino exchange & Heavy neutrino exchange \tabularnewline
\hline 
\T
$^{48}$Ca  &1.98(59)		&16.3(95)		  \tabularnewline
$^{76}$Ge  &5.42(103)		&48.1(255)		   	 \tabularnewline
$^{82}$Se  &4.37(83)		&35.6(189)		   	 \tabularnewline
$^{96}$Zr	&2.53(40)		&59.0(309)		   	 \tabularnewline
$^{100}$Mo &3.73(60)		&99.3(516)	   	  \tabularnewline
$^{110}$Pd &3.62(58)		&95.7(498)		   	 \tabularnewline
$^{116}$Cd &2.78(44)		&67.1(321)		   	 \tabularnewline
$^{124}$Sn &3.50(67)		&37.8(200)		   	 \tabularnewline
$^{128}$Te &4.48(85)		&48.4(257)		   	 \tabularnewline
$^{130}$Te &4.03(77)		&44.0(233)		   	 \tabularnewline
$^{136}$Xe &3.33(63)		&35.1(186)	   	  \tabularnewline
$^{148}$Nd &1.98(32)		&59.4(309)		   	 \tabularnewline
$^{150}$Nd &2.32(37)		&68.4(356)		   	 \tabularnewline
$^{154}$Sm &2.50(40)		&67.1(349)		   	 \tabularnewline
$^{160}$Gd &3.62(58)		&92.9(483)		   	 \tabularnewline
$^{198}$Pt &1.88(30)		&61.5(320)		   	 \tabularnewline
\end{tabular}\end{ruledtabular} 
\end{table}

\begin{figure}[h]
\begin{center}
\includegraphics[width=8.6cm]{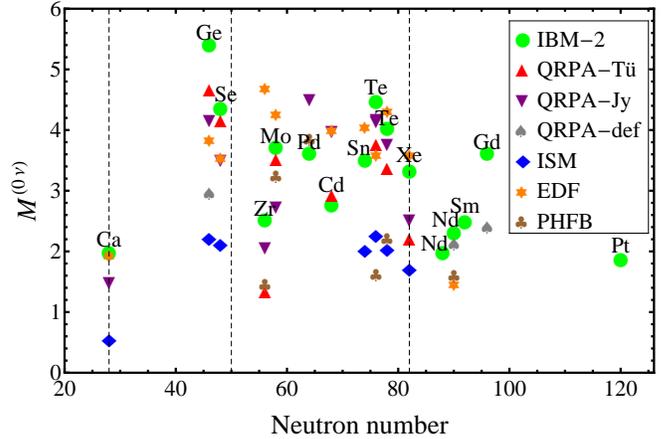} 
\end{center}
\caption{\label{allnme}(Color online) IBM-2 results for $0\nu\beta\beta$ nuclear matrix elements compared with QRPA-T\"u \cite{simkovic1}, the ISM \cite{poves}, QRPA-Jy \cite{suh,suh07,toi95,kor07}, QRPA-deformed \cite{fan11}, DFT \cite{Martinez-Pinedo}, and HFB \cite{chandra}.}
\end{figure}

Finally, having investigated the effect of short-range correlations on $0\nu\beta\beta$ we are now able to compare our results with all available calculations done with the same SRC including DFT \cite{Martinez-Pinedo} and HFB \cite{chandra}. These are shown in Fig.~\ref{allnme}. We note now that while the ISM/QRPA/IBM-2 have the same trend with $A$, the other two do not. For the isotopic ratio $M^{(0\nu)}(^{128}$Te)/$M^{(0\nu)}(^{130}$Te) the DFT method gives 0.86 in sharp contrast with the value 1.11. Also, while the ISM/QRPA/IBM-2 have a small value for $^{96}$Zr, DFT has a large value. We therefore conclude that the approximations made in the DFT/HFB lead to a different behavior with $A$. This point is currently being investigated \cite{rodr}. Also, the Fermi matrix elements in the DFT are comparable to those in IBM-2 and larger than those in the ISM \cite{rodr}.

\subsection{Limits on neutrino mass}
\label{limits}
\subsubsection{Light neutrino exchange}

\begin{figure}[h]
\begin{center}
\includegraphics[width=8.6cm]{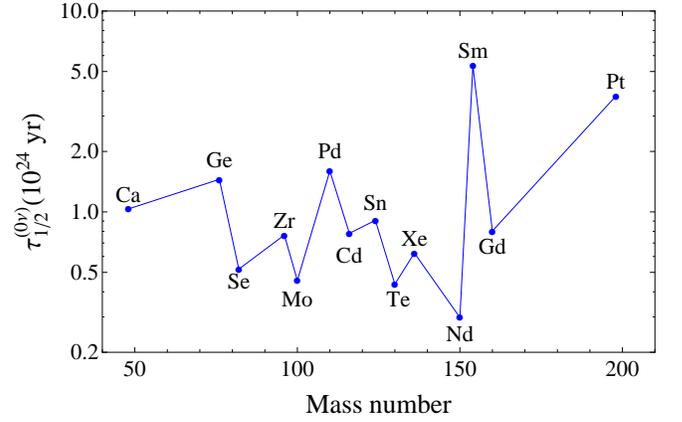} 
\end{center}
\caption{\label{fig6}(Color online) Expected half-lives for $\left\langle m_{\nu}\right\rangle=1$~eV, $g_{A}=1.269$. The points for $^{128}$Te and $^{148}$Nd decays are not included in this figure. The figure is in semilogarithmic scale.}
\end{figure}

The calculation of nuclear matrix elements in IBM-2 can now be combined
with the phase-space factors calculated in \cite{kotila} and given
in Table III and Fig. 8 of that reference to produce our final results
for half-lives for light neutrino exchange in Table~\ref{table10} and Fig.~\ref{fig6}. The half-lives are calculated using the formula 
\begin{equation}
\lbrack\tau_{1/2}^{0\nu}]^{-1}=G_{0\nu}^{(0)}\left\vert M_{0\nu}\right\vert ^{2}\left\vert \frac{\left\langle m_{\nu}\right\rangle }{m_{e}}\right\vert ^{2}.
\end{equation}
We note here that the combination must be done consistently.
If the value of $g_A$ is included in $M_{0\nu}$, then it should not be included in $G_{0\nu}^{(0)}$ and similarly for a factor of 4 included in some definition of $G_{0\nu}^{(0)}$ \cite{tomoda} and not in others \cite{boehm}. See Eq.~(53) of Ref.~\cite{kotila}. This point has caused considerable confusion in the literature.
In Table~\ref{table10} and Fig.~\ref{fig6} the values $\left\langle m_{\nu}\right\rangle =1$~eV
and $g_{A}=1.269$ are used. For other values they can be scaled with
$\left\vert \left\langle m_{\nu}\right\rangle /m_{e}\right\vert ^{2}$
and $g_{A}^{4}$.
\begin{ruledtabular}
\begin{table}[h]
\caption{\label{table10}Left: Calculated half-lives in IBM-2 M-S SRC for neutrinoless double-$\beta$ decay for $\left<m_{\nu}\right>=1$~eV and $g_A=1.269$. Right: Upper limit on neutrino mass from current experimental limit from a compilation of Barabash \cite{barabash11}. The values reported by Klapdor-Kleingrothaus \textit{et al.} \cite{klapdor}, the IGEX Collaboration \cite{igex}, and the recent limits from KamLAND-Zen \cite{kamland} and EXO \cite{exo0nu} are also included.}
\begin{tabular}{lc|cc}
Decay  &  \ensuremath{\tau_{1/2}^{0\nu}}(\ensuremath{10^{24}}yr) &  \ensuremath{\tau_{1/2, exp}^{0\nu}}(yr) &$\left< m_{\nu}\right>$ (eV)\\
 \hline
 \T
$^{48}$Ca$\rightarrow ^{48}$Ti		&1.03 &$>5.8\times 10^{22}$ &$<4.2$\\
$^{76}$Ge$\rightarrow ^{76}$Se	 	&1.45 &$>1.9\times 10^{25}$ &$<0.28$\\
								 	&	 	&$1.2\times 10^{25}$\footnotemark[1] &$0.35$\\
								 	&	 	&$>1.6\times 10^{25}$\footnotemark[2] &$<0.30$\\
$^{82}$Se$\rightarrow ^{82}$Kr	 	&0.52 &$>3.6\times 10^{23}$ &$<1.2$\\
$^{96}$Zr$\rightarrow ^{96}$Mo		&0.77 &$>9.2\times 10^{21}$ &$<9.1$\\
$^{100}$Mo$\rightarrow ^{100}$Ru 	&0.46 &$>1.1\times 10^{24}$ &$<0.64$\\
$^{110}$Pd$\rightarrow ^{110}$Cd 	&1.60 & &\\
$^{116}$Cd$\rightarrow ^{116}$Sn  	&0.78 &$>1.7\times 10^{23}$ &$<2.1$\\
$^{124}$Sn$\rightarrow ^{124}$Te 	&0.91 & &\\
$^{128}$Te$\rightarrow ^{128}$Xe 	&8.53 &$>1.5\times 10^{24}$ &$<2.4$\\
$^{130}$Te$\rightarrow ^{130}$Xe 	&0.44 &$>2.8\times 10^{24}$ &$<0.39$\\
$^{136}$Xe$\rightarrow ^{136}$Ba 	&0.62 &$>5.7\times 10^{24}$\footnotemark[3] &$<0.33$\\
								 	&	 &$>1.6\times 10^{25}$\footnotemark[4] &$<0.20$\\

$^{148}$Nd$\rightarrow ^{148}$Sm 	&2.54 & &\\
$^{150}$Nd$\rightarrow ^{150}$Sm 	&0.30 &$>1.8\times 10^{22}$ &$<4.1$\\
$^{154}$Sm$\rightarrow ^{154}$Gd 	&5.34 & &\\
$^{160}$Gd$\rightarrow ^{160}$Dy 	&0.80 & &\\
$^{198}$Pt$\rightarrow ^{198}$Hg 	&3.77 & &\\
\end{tabular}
\footnotetext[1]{Reference~\cite{klapdor}.}
\footnotetext[2]{Reference~\cite{igex}.}
\footnotetext[3]{Reference~\cite{kamland}.}
\footnotetext[4]{Reference~\cite{exo0nu}.}

\end{table}
\end{ruledtabular}

\begin{figure}[h]
\begin{center}
\includegraphics[width=8.6cm]{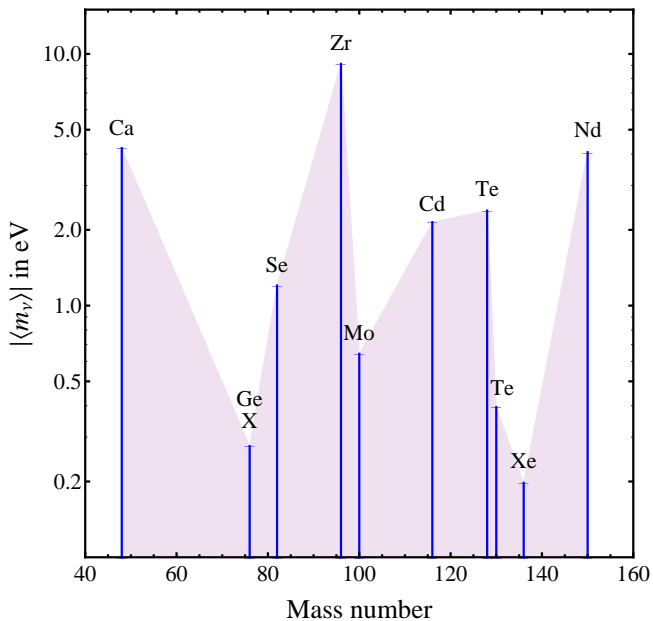} 
\end{center}
\caption{\label{fig7}(Color online) Limits on neutrino mass from current experimental limits from a compilation of A.\ Barabash \cite{barabash11} and recent measurement for $^{136}$Xe from EXO \cite{exo0nu}. The values reported by Klapdor-Kleingrothaus \textit{et al.} \cite{klapdor} is shown by the symbol X. The figure is in semilogarithmic scale. The shaded area represents the values of $|\langle m_{\nu}\rangle|$ allowed by the current experiments.}
\end{figure}

The effective neutrino mass is the quantity we want to extract from
experiment. Unfortunately, the axial vector coupling constant is renormalized
in nuclei to $g_{A,eff}$. A (model-dependent) estimate of $g_{A,eff}$
can be obtained from the experimental knowledge of single-$\beta$ decay
and/or of 2$\nu\beta\beta$ decay. This will be discussed in the following
section. Here we show in Fig.~\ref{fig7} and Table~\ref{table10}, the limits on neutrino mass from current
experimental upper limits using IBM-2 matrix elements of Table~\ref{table5} and $g_{A}=1.269$. In addition to the experimental upper limits, a value has been reported for the half-life in $^{76}$Ge, $1.2\times 10^{25}$yr \cite{klapdor}. This is also reported in Fig.~\ref{fig7}. 
\begin{figure}[cbt!]
\begin{center}
\includegraphics[width=8.6cm]{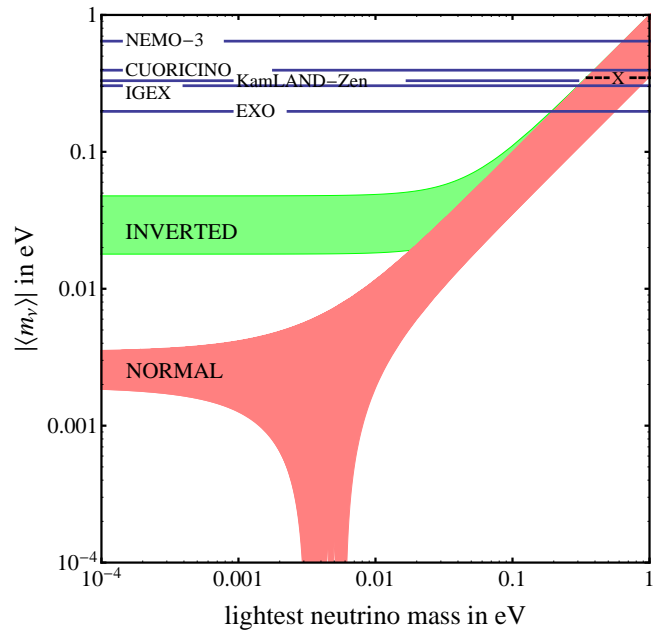} 
\end{center}
\caption{\label{fig8}(Color online) Current limits to $\left\langle m_{\nu}\right\rangle$ from CUORICINO \cite{cuoricino}, IGEX \cite{igex}, NEMO-3 \cite{nemo}, KamLAND-Zen \cite{kamland}, and EXO \cite{exo0nu}, and IBM-2 M-S SRC nuclear matrix elements. The value of Ref.~\cite{klapdor} is shown by an X. It is consistent only with nearly degenerate neutrino masses.  The figure is in logarithmic scale.}
\end{figure}

The average light neutrino mass is constrained by atmospheric, solar,
reactor and accelerator neutrino oscillation experiments to be \cite{fogli}
\begin{equation}
\begin{split}
\left\langle m_{\nu}\right\rangle   =  \left\vert c_{13}^{2}c_{12}^{2}m_{1}\right.&\left.+c_{13}^{2}s_{12}^{2}m_{2}e^{i\varphi_{2}}+s_{13}^{2}m_{3}e^{i\varphi_{3}}\right\vert , \\
c_{ij}  =  \cos\vartheta_{ij},\text{ \ \ }&s_{ij}=\sin\vartheta_{ij},\text{ \ \ }\varphi_{2,3}=[0,2\pi],\\
\left(m_{1}^{2},m_{2}^{2},m_{3}^{2}\right)  =&  \frac{m_{1}^{2}+m_{2}^{2}}{2}\\
&+\left(-\frac{\delta m^{2}}{2},+\frac{\delta m^{2}}{2},\pm\Delta m^{2}\right).
\end{split}
\end{equation}
Using the best fit values \cite{fogli} 
\begin{equation}
\begin{split}
\sin^{2}\vartheta_{12} &=0.213,\text{\ \ }\sin^{2}\vartheta_{13}=0.016,\\
\sin^{2}\vartheta_{23}&=0.466,\text{\ \ }\delta m^{2} =7.67\times10^{-5}\text{~eV}^{2},\\
\Delta m^{2}&=2.39\times10^{-3}\text{~eV}^{2}
\end{split}
\end{equation}
we obtain the values given in Fig.~\ref{fig8}. In this figure we have added
the current limits, for $g_{A}=1.269$, coming from CUORICINO \cite{cuoricino}, IGEX \cite{igex}, NEMO-3 \cite{nemo}, KamLAND-Zen \cite{kamland}, and the EXO \cite{exo0nu} experiment. Also, henceworth we use 
 $c=1$ to conform with standard notation.

\subsubsection{Heavy neutrino exchange}

\begin{figure}[cbt!]
\begin{center}
\includegraphics[width=8.6cm]{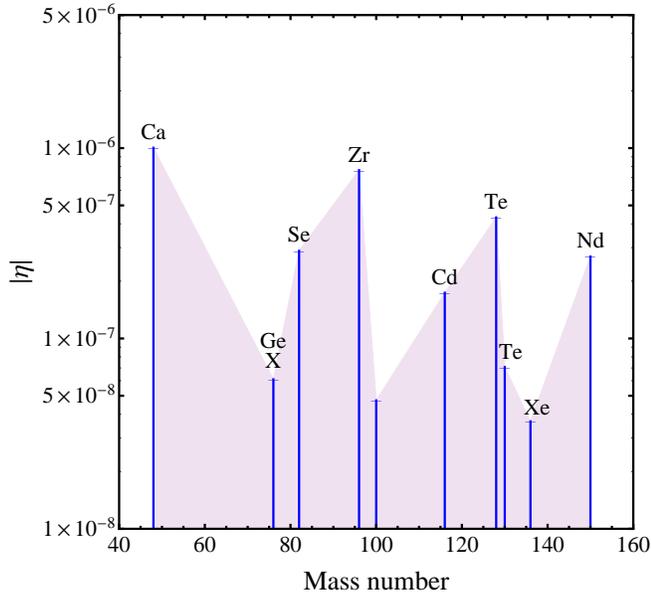} 
\end{center}
\caption{\label{fig9}(Color online) Limits on the lepton nonconserving parameter $|\eta|$. The value of Ref.~\cite{klapdor} is shown by an X. The figure is in semilogarithmic scale. The shaded area represents the values of $|\eta|$ allowed by the current experiments.}
\end{figure}

The half-lives for this case are calculated using the formula 
\begin{equation}
\begin{split}
\lbrack\tau_{1/2}^{0\nu_{h}}]^{-1} & =  G_{0\nu}^{(0)}\left\vert M_{0\nu_{h}}\right\vert ^{2}\left\vert \eta\right\vert ^{2}\\\
\eta & \equiv  m_{p}\left\langle m_{\nu_{h}}^{-1}\right\rangle =\sum_{k=heavy}\left(U_{ek_h}\right)^{2}\frac{m_{p}}{m_{k_h}}.
\end{split}
\end{equation}
The expected half-lives for $|\eta|=2.75\times10^{-7}$, and using the IBM-2 matrix elements of Table~\ref{table7},
are shown in Table~\ref{table11}. For other values of $\eta$ they scale as $|\eta|^2$.
There are no direct experimental bounds on $\eta $.
Recently, Tello\textit{\ et al.} \cite{tello} have argued that from
lepton flavor violating processes and from Large Hadron Collider (LHC) experiments one can
put some bounds on the right-handed leptonic mixing matrix $U_{ek,heavy}$
and thus on $\eta$. In the model of Ref.~\cite{tello}, when converted to our notation, $\eta$ can be written as 
\begin{equation}
\eta=\frac{M_W^4}{M_{WR}^4}\sum_{k=heavy}\left( V_{ek_h}\right)^2\frac{m_p}{m_{k_h}},
\end{equation}
where $M_W$ is the mass of the $W$-boson, $M_W=(80.41\pm 0.10)$~GeV \cite{experiment}, $M_{WR}$ is the mass of the $WR$-boson, assumed in \cite{tello} to be $M_{WR}=3.5$~TeV and $V=\left( M_{WR}/M_W \right)^2 U$. The ratio $\left( M_W / M_{WR} \right)^4$ is then $2.75 \times 10^{-7}$, the value we have used in the left portion of Table~\ref{table11}. By comparing the calculated half-lives with their current experimental limits, we can set limits on the lepton-nonconserving parameter $|\eta|$, shown in Table~\ref{table11} and Fig.~\ref{fig9}.
\begin{ruledtabular}
\begin{center}
\begin{table*}[cbt!]
\caption{\label{table11}Left: Calculated half-lives for neutrinoless double-$\beta$ decay with exchange of heavy neutrinos for $\eta=2.75\times10^{-7}$ and $g_A=1.269$. Right: Upper limits of $|\eta|$ and lower limits of heavy neutrino mass (see text for details) from current experimental limit from a compilation of Barabash \cite{barabash11}. The value reported by Klapdor-Kleingrothaus \textit{et al.} \cite{klapdor}, and the IGEX Collaboration \cite{igex}, and the recent limits from KamLAND-Zen \cite{kamland} and EXO \cite{exo0nu} are also included.}
\begin{tabular}{lc|ccc}
Decay  &  \ensuremath{\tau_{1/2}^{0\nu_h}}(\ensuremath{10^{24}}yr) &  \ensuremath{\tau_{1/2, exp}^{0\nu_h}}(yr) &$|\eta|(10^{-6})$ &$\left< m_{\nu_h}\right>$(GeV)\\
 \hline
 \T
$^{48}$Ca$\rightarrow ^{48}$Ti		&0.77	&$>5.8\times 10^{22}$				&<1.00		&>0.26\\
$^{76}$Ge$\rightarrow ^{76}$Se	 	&0.95	&$>1.9\times 10^{25}$				&<0.061		&>4.2\\
								 	&	 	&$1.2\times 10^{25}$\footnotemark[1]&0.077 	&3.362\\
								 	&	 	&$>1.6\times 10^{25}$\footnotemark[2]&<0.066 	&>3.88\\
$^{82}$Se$\rightarrow ^{82}$Kr	 	&0.40	&$>3.6\times 10^{23}$				&<0.29		&>0.89\\
$^{96}$Zr$\rightarrow ^{96}$Mo		&0.07	&$>9.2\times 10^{21}$				&<0.77		&>0.34\\
$^{100}$Mo$\rightarrow ^{100}$Ru 	&0.03	&$>1.1\times 10^{24}$				&<0.0047	&>5.5\\
$^{110}$Pd$\rightarrow ^{110}$Cd 	&0.12	&									&			&\\
$^{116}$Cd$\rightarrow ^{116}$Sn  	&0.07	&$>1.7\times 10^{23}$				&<0.17		&>1.5\\
$^{124}$Sn$\rightarrow ^{124}$Te 	&0.39	&									&			&\\
$^{128}$Te$\rightarrow ^{128}$Xe 	&3.71	&$>1.5\times 10^{24}$				&<0.43		&>0.60\\
$^{130}$Te$\rightarrow ^{130}$Xe 	&0.19	&$>2.8\times 10^{24}$				&<0.071		&>3.6\\
$^{136}$Xe$\rightarrow ^{136}$Ba 	&0.29	&$>5.7\times 10^{24}$\footnotemark[3]&<0.061	&>4.2\\
								 	&		&$>1.6\times 10^{25}$\footnotemark[4]&<0.116	&>2.2\\
$^{148}$Nd$\rightarrow ^{148}$Sm 	&0.14	&									&			&\\
$^{150}$Nd$\rightarrow ^{150}$Sm 	&0.02	&$>1.8\times 10^{22}$				&<0.27		&>0.96\\
$^{154}$Sm$\rightarrow ^{154}$Gd 	&0.38	&									&			&\\
$^{160}$Gd$\rightarrow ^{160}$Dy 	&0.06	&									&			&\\
$^{198}$Pt$\rightarrow ^{198}$Hg 	&0.18	&									&			&\\
\end{tabular}
\footnotetext[1]{Reference~\cite{klapdor}.}
\footnotetext[2]{Reference~\cite{igex}.}
\footnotetext[3]{Reference~\cite{kamland}.}
\footnotetext[4]{Reference~\cite{exo0nu}.}
\end{table*}
\end{center}
\end{ruledtabular}
If we write
\begin{equation}
\eta=\frac{M_W^4}{M_{WR}^4} \frac{m_p}{ \langle m_{\nu_h} \rangle },
\end{equation}
we can also set limits on the average heavy neutrino mass, $\langle m_{\nu_h} \rangle$, as shown in the last column of Table~\ref{table11}. This limit is model dependent since it is assumed that $M_{WR}=3.5$~TeV. For other values of $M_{WR}$ it scales as $M_{WR}^{-4}$.

If both light and heavy neutrino exchange contribute, the half-lives are given by 
\begin{equation}
\lbrack\tau_{1/2}^{0\nu}]^{-1} = G_{0\nu}^{(0)}\left\vert M_{0\nu} \frac{\langle m_{\nu}\rangle }{m_e}  +  M_{0\nu_{h}} \eta \right\vert ^{2}.
\end{equation}
It is interesting  to note here the possibility of interference between light and heavy neutrino exchange, as emphasized recently by several authors. The limits presented in Figs.~\ref{fig7}, \ref{fig8},and \ref{fig9} are based on the calculation with M-S SRC. If CCM SRC are used, they should be multiplied by $\sim 1.2$ (light neutrino exchange) and $\sim 2.0$ (heavy neutrino exchange).

\section{2$\nu\beta\beta$ decay}
\label{sect2nu}
\subsection{Matrix elements}
As mentioned in the previous section, the calculated matrix elements of the GT operator in single-$\beta$ decay appear to be systematically larger than those extracted from the measured $ft$ values of allowed GT transitions. To take into account these results, it has been found convenient to renormalize the value of $g_A$ to be used in a particular model calculation by introducing an effective $g_{A,eff,\beta}$ defined as
\begin{equation}
\left( \frac{g_{A,eff,\beta}}{g_A}\right)=\frac{\left|M_{exp,\beta} \right|}{\left|M_{th,\beta} \right|},
\end{equation}
where $g_A=1.269$ and $M_{\beta}$ are the matrix elements for single-$\beta$ decay. The ratio $(g_{A,eff}/g_A)$ is also called the quenching, $q$, or hindrance, $h=1/q$, factor. The quenching of $g_A$ comes from two effects: (i) the limited model space in which the calculation is done and (ii) the contribution of non-nucleonic degrees of freedom, $\Delta$,.... The first type of quenching depends strongly on the size of the model space used in the calculation and is thus model dependent. It was extensively investigated in light nuclei, $A\sim20$, in the 1970s \cite{wil73a,wil73b,wil74} within the framework of the ISM where it was found that $g_{A,eff}\cong 1.0$, $q\cong 0.75$. In heavy nuclei, of particular interest in this paper, the question of quenching was first discussed by Fujita and Ikeda \cite{fuj65} in 1965. These authors analyzed $\beta$-decay in mass $A\sim120$ nuclei within the framework of various models (pairing, pairing plus quadrupole, etc.) and found very small quenching factors, $q\simeq 0.2-0.3$ thus stimulating the statement that massive renormalization of $g_A$ occurs in heavy nuclei \cite{wil74}. The second type of quenching was extensively investigated theoretically in the 1970s \cite{eri71,bar74,ari73}. This effect does not depend much on the nuclear model used in the calculation, but rather on the mechanism of coupling to non-nucleonic degrees of freedom. It is being re-investigated currently within the framework of chiral effective field theory (EFT) \cite{men11} and there are hints that it has a complex structure, in particular that it may depend on momentum transfer and that it may lead in some cases to an enhancement rather than a quenching.

The values of $g_{A,eff}$ depend crucially on the model used through the size and composition of the model space, especially on whether or not spin-orbit partners are included in the calculation. For example, while the QRPA includes spin-orbit partners, IBM-2 and the ISM do not. Conversely, in the ISM calculations the size of the model space is $\sim10^9$, while in the QRPA and IBM-2 it is much smaller. In order to extract $g_{A,eff,\beta}$ for a given mass number $A$, one needs to do a calculation of single-$\beta$ decay in that region and compare with experiment where available. Within the context of IBM-2, some calculations were done in the 1980s \cite{del89}. Very recently, the problem has been readdressed and results will be published soon \cite{yos12}.

Double-$\beta$ decay depends on $g_A$ as $g_A^4$ and thus its quenching is of extreme importance. Since $2\nu\beta\beta$ decay has now been measured in several nuclei, it provides another way to estimate $g_{A,eff}$ which we denote by $g_{A,eff,2\nu\beta\beta}$. In this section, we attempt an estimation of $g_{A,eff,2\nu\beta\beta}^{\rm{IBM-2}}$ within the framework of IBM-2 in the closure approximation and also extract $g_{A,eff,2\nu\beta\beta}^{\rm{ISM}}$ within the framework of the ISM in the non-closure approximation. The extraction of $g_{A,eff,2\nu\beta\beta}^{\rm{IBM-2}}$ in the non-closure approximation will be presented in the forthcoming publication mentioned above \cite{yos12}.

2$\nu\beta\beta$ is a process allowed by the standard model
and thus in principle exactly calculable. The theory of 2$\nu\beta\beta$
decay was developed by Primakoff and Rosen \cite{primakoff}, Konopinski
\cite{konopinski}, Doi \textit{et al.} \cite{doi} and Haxton \textit{et
al.} \cite{haxton}. The calculation of $2\nu\beta\beta$ turns out
to be more complex than that of $0\nu\beta\beta$. 

\begin{figure}[cbt!]
\begin{center}
\includegraphics[width=8.6cm]{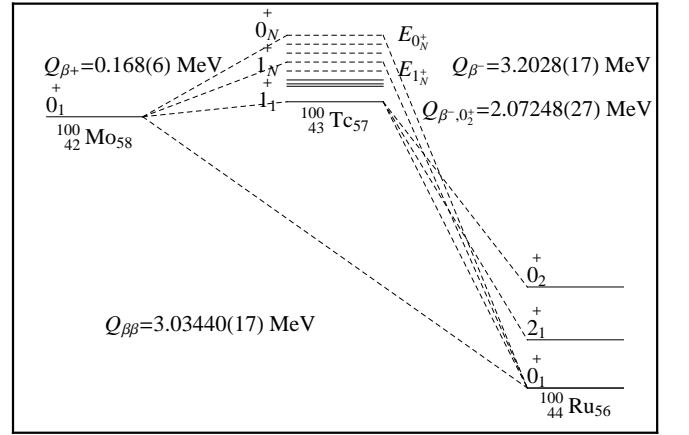} 
\end{center}
\caption{\label{fig10}The decay $_{42}^{100}$Mo$_{58}  \rightarrow   _{44}^{100}$Ru$_{56}$, an example of $2\nu\beta\beta$ decay.}
\end{figure}

(i) The closure
approximation may not be good and one needs to evaluate explicitly
the matrix elements to and from the individual $1_{N}^{+}$ and $0_{N}^{+}$
states in the intermediate odd-odd nucleus, Fig.~\ref{fig10}, 
\begin{equation}
M_{GT,N}^{(2\nu)}=\frac{\left\langle 0_{F}^{+}\left\Vert \tau^{\dag}\vec{\sigma}\right\Vert 1_{N}^{+}\right\rangle \left\langle 1_{N}^{+}\left\Vert \tau^{\dag}\vec{\sigma}\right\Vert 0_{I}^{+}\right\rangle }{\frac{1}{2}\left(Q_{\beta\beta}+2m_{e}c^{2}\right)+E_{1_{N}^{+}}-E_{I}},
\end{equation}
and 
\begin{equation}
M_{F,N}^{(2\nu)}=\frac{\left\langle 0_{F}^{+}\left\Vert \tau^{\dag}\right\Vert 0_{N}^{+}\right\rangle \left\langle 0_{N}^{+}\left\Vert \tau^{\dag}\right\Vert 0_{I}^{+}\right\rangle }{\frac{1}{2}\left(Q_{\beta\beta}+2m_{e}c^{2}\right)+E_{0_{N}^{+}}-E_{I}}.
\end{equation}
This evaluation has been done in selected nuclei within the framework
of the pnQRPA \cite{QRPAnoclosure}, the proton-neutron microscopic anharmonic vibrator approach (pnMAVA) \cite{kotila09,kotila10}, and the ISM \cite{ISMnoclosure, caurier2007,horoi} and it has been programmed very recently within the framework of the proton-neutron interacting boson-fermion model (IBFM-2)
\cite{yos12}. The calculation requires the difficult task of determining
the structure of the intermediate odd-odd nucleus.

(ii) The PSFs cannot be exactly separated \ from the nuclear
matrix elements. These factors must be calculated separately for each
state $1_{N}^{+}$ and $0_{N}^{+}$. In order to calculate half-lives
and other observable quantities the product of the PSFs $G_{2\nu,N}^{(i)}$
and matrix elements must be calculated and the contributions summed
over all individual states. This is a daunting problem compounded
by the fact that in most calculations the giant Gamow-Teller resonance
that contributes to the matrix elements is not included in the model
space.

The separation of PSFs and nuclear matrix elements can be done in two
cases: (1) the closure approximation (CA) and (2) the single-state
dominance (SSD) approximation \cite{SSD1,SSD2,civitarese,domin1,domin2}. \ In both cases, the inverse half-life
can be written as
\begin{equation}
\left[\tau_{1/2}^{2\nu}\right]^{-1}=G_{2\nu}^{(0)}\left\vert m_{e}c^{2}M_{2\nu}\right\vert ^{2}.
\end{equation}

In the CA, the matrix elements $M_{2\nu}$
can be written as 
\begin{equation}
\begin{split}
M_{2\nu} & =  g_{A}^{2}M^{(2\nu)}, \\
M^{(2\nu)} & =  -\left[\frac{M_{GT}^{(2\nu)}}{\tilde{A}_{GT}}-\left(\frac{g_{V}}{g_{A}}\right)^{2}\frac{M_{F}^{(2\nu)}}{\tilde{A}_{F}}\right],
\end{split}
\end{equation}
where 
\begin{equation}
\begin{split}
M_{GT}^{(2\nu)} & =  \left\langle 0_{F}^{+}\left\vert \sum_{nn^{\prime}}\tau_{n}^{\dag}\tau_{n^{\prime}}^{\dag}\vec{\sigma}_{n}\cdot\vec{\sigma}_{n^{\prime}}\right\vert 0_{I}^{+}\right\rangle , \\
M_{F}^{(2\nu)} & =  \left\langle 0_{F}^{+}\left\vert \sum_{nn^{\prime}}\tau_{n}^{\dag}\tau_{n^{\prime}}^{\dag}\right\vert 0_{I}^{+}\right\rangle .
\end{split}
\end{equation}
The closure energies $\tilde{A}_{GT}$ and $\tilde{A}_{F}$ are defined
by
\begin{equation}
\begin{split}
\tilde{A}_{GT} & =  \frac{1}{2}\left(Q_{\beta\beta}+2m_{e}c^{2}\right)+\left\langle E_{1^{+},N}\right\rangle -E_{I}, \\
\tilde{A}_{F} & =  \frac{1}{2}\left(Q_{\beta\beta}+2m_{e}c^{2}\right)+\left\langle E_{0^{+},N}\right\rangle -E_{I},
\end{split}
\end{equation}
where $\left\langle E_{N}\right\rangle $ is a suitable chosen excitation
energy in the intermediate odd-odd nucleus. The Fermi matrix elements
are suppressed by isospin considerations and are often neglected.
In case they are not, care must be taken since the closure energy $\tilde{A}_{F}$ is different from the closure energy $\tilde{A}_{GT}$, although, for simplicity $\tilde{A}_{GT}=\tilde{A}_{F}=\tilde{A}$
is often used.

In the SSD approximation, the matrix elements
are given by 
\begin{equation}
\begin{split}
M_{GT,SSD}^{(2\nu)} & =  \frac{\left\langle 0_{F}^{+}\left\Vert \tau^{\dag}\vec{\sigma}\right\Vert 1_{1}^{+}\right\rangle \left\langle 1_{1}^{+}\left\Vert \tau^{\dag}\vec{\sigma}\right\Vert 0_{I}^{+}\right\rangle }{\frac{1}{2}\left(Q_{\beta\beta}+2m_{e}c^{2}\right)+E_{1_{1}^{+}}-E_{I}}, \\
M_{F,SSD}^{(2\nu)} & =  \frac{\left\langle 0_{F}^{+}\left\Vert \tau^{\dag}\right\Vert 0_{1}^{+}\right\rangle \left\langle 0_{1}^{+}\left\Vert \tau^{\dag}\right\Vert 0_{I}^{+}\right\rangle }{\frac{1}{2}\left(Q_{\beta\beta}+2m_{e}c^{2}\right)+E_{0_{1}^{+}}-E_{I}},
\end{split}
\end{equation}
where $E_{1_{1}^{+}}$ and $E_{0_{1}^{+}}$ are the energies in the
intermediate odd-odd nucleus of the single state that dominates the
decay. From these one can form the quantities
\begin{equation}
\begin{split}
M_{SSD}^{(2\nu)} & =  -\left[M_{GT,SSD}^{(2\nu)}-\left(\frac{g_{V}}{g_{A}}\right)^{2}M_{F,SSD}^{(2\nu)}\right],\\
M_{2\nu,SSD} & =  g_{A}^{2}M_{SSD}^{(2\nu)},
\end{split}
\end{equation}
and calculate the half-lives from (25), with $G_{2\nu}^{(0)}$ given
by $G_{2\nu,SSD}^{(0)}$.

\subsection{Results}

In this article, we present results of a calculation of the nuclear
matrix elements for 2$\nu\beta\beta$ in the CA using the transition operator of Sec. II. A. In this case only
the terms $\tilde{h}_{VV}^{F}$ and $\tilde{h}_{AA}^{GT}$ are considered.
An advantage of the closure approximation for 2$\nu\beta\beta$ decay
is that the nuclear matrix elements can be calculated using the same
method discussed in Sec. II, by simply replacing the neutrino potential
$v(p)$ by 
\begin{equation}
v_{2\nu}(p)=\frac{\delta(p)}{p^{2}},
\end{equation}
which is the Fourier-Bessel transform of the configuration space potential
$V(r)=1$. Since our purpose here is a direct comparison of 2$\nu\beta\beta$
and 0$\nu\beta\beta$ decays, this avoids possible systematic and
accidental errors. The CA is not expected to be good  for $2\nu\beta\beta$ decay, since only $1^+$ and $0^+$ intermediate states in the odd-odd nucleus contribute to the decay. We use it here only as an \textit{estimate}, with appropriately chosen closure energy \cite[p.71]{tomoda} in order to extract $g_{A,eff,2\nu\beta\beta}\equiv g_{A,eff}$, which is the purpose of this section.

Our calculated matrix elements for $2\nu\beta\beta$ decay are shown in Table~\ref{table12}. Also here as in Table~\ref{table3} we have separated the class of nuclei where protons and neutrons occupy the same major shell from those where they do not, and we added $A=48$ at the bottom of the table. The problem of spuriosity of the Fermi matrix elements is here even more acute than in the case of $0\nu\beta\beta$. In the absence of isospin violation, all Fermi matrix elements for $2\nu\beta\beta$ should be exactly zero. Breaking of isospin is present in all calculations (ISM, QRPA, IBM-2, DFT, and HFB). Within the model space and in the closure approximation that we are using, we have a large breaking for nuclei in which protons and neutrons occupy the same major shell  and zero breaking in the others. The small value $\sim0.02$ in Table~\ref{table12} is an indication of our numerical accuracy in calculating overlap of wave functions. From the dimensionless matrix elements in Table~\ref{table12} and the values of $\tilde{A}$ we calculate the values of $|m_e c^2 M^{(2\nu)}|$ given in Table~\ref{table13}. In constructing this table we have taken into account only the GT matrix elements, since, as mentioned above, the IBM-2 F matrix elements are largely spurious in nuclei where protons and neutrons occupy the same major shell.
\begin{table}[cbt!]
 \caption{\label{table12}$2\nu\beta\beta$ matrix elements (dimensionless) to the ground state
(columns 2 and 3) and to the first excited state (columns 4 and 5)
using the microscopic interacting boson model (IBM-2) in the closure
approximation.}
 \begin{ruledtabular} %
\begin{tabular}{ccccc}
$A$ &\multicolumn{2}{c}{$0_{1}^{+}$} &\multicolumn{2}{c}{$0_{2}^{+}$}\\ \cline{2-3} \cline{4-5}
\T
  & $M_{GT}^{(2\nu)}$  & $M_{F}^{(2\nu)}$  & $M_{F}^{(2\nu)}$  & $M_{GT}^{(2\nu)}$ \tabularnewline
\hline 
\T
76  & 4.34  &-2.69   & -1.35  & 1.99\tabularnewline
82  & 3.50  &-2.39   & -0.83  & 1.03\tabularnewline
\hline
96  & 2.22  &0.02    & 0.00  & 0.04\tabularnewline
100  &2.94  & 0.03   & 0.00  & 0.39\tabularnewline
110  & 2.98  &0.03   & 0.01  & 1.48\tabularnewline
116  & 2.31  &0.02   & 0.01  & 0.86\tabularnewline
\hline
124  & 2.80  &-1.60   & -1.34  & 2.15\tabularnewline
128  & 3.63  &-1.99   & -1.53  & 2.65\tabularnewline
130  & 3.31  &-1.79   & -1.45  & 2.59\tabularnewline
136  & 2.76  &-1.44   & -0.83  & 1.63\tabularnewline
\hline
148  & 1.24  &0.02   & 0.00  & 0.17\tabularnewline
150  & 1.54  &0.02   & 0.00  & 0.29\tabularnewline
154  & 1.91  &0.02   & -0.00  & 0.05\tabularnewline
160  & 2.99  &0.02  & 0.01  & 0.51\tabularnewline
198  & 1.00  &0.01   & 0.00  & 0.03\tabularnewline
\hline
48  & 1.57  &-1.08   & -4.77  & 5.02\tabularnewline
\end{tabular}
\end{ruledtabular} 
\end{table}

We  investigate two choices of $\tilde{A}_{GT}$.
The first choice is that taken from  Ref.~\cite{haxton} or estimated by the systematics, $\tilde{A}_{GT}=1.12A^{1/2}$~MeV, where $A$ without tilde denotes the mass number. In cases where
transitions between spin-orbit partners dominate, one expects the SSD approximation to be appropriate.
Our second choice is SSD for  $_{40}$Zr, $_{42}$Mo, $_{46}$Pd, and $_{48}$Cd, where the dominant transition is $g_{9/2}$-$g_{7/2}$, 
and $_{60}$Nd where the dominant transition is $h_{11/2}$-$h_{9/2}$. 
 In the same table we also show the values of the matrix elements in the ISM without the closure approximation \cite{caurier2007}.
The ISM calculation are all in nuclei in which protons and neutrons occupy the same major shell. By comparing these calculations with those in IBM-2 with the Fermi matrix elements set to zero we see that the two calculations have the same behavior with mass number but differ by a factor of approximately 2.  The last columns in Table~\ref{table13} gives the values of the matrix elements $|M_{2\nu}^{eff}|$ extracted from experiment \cite{kotila}.
\begin{ruledtabular}
\begin{center}
\begin{table*}[ctb!]
\caption{\label{table13}Calculated values of $2\nu\beta\beta$ matrix elements in IBM-2 with $g_A=1.269$ and the ISM with $g_A=1.25$.}
\begin{tabular}{cccccccc}
$A$ &\multicolumn{2}{c}{ \ensuremath{\tilde{A}}(MeV)}  &\multicolumn{5}{c}{  \ensuremath{\left\vert m_{e}c^{2}M^{(2\nu)}\right\vert}}\\ \cline{2-3} \cline{4-8}
\T
&\multirow{2}{*}{$\tilde{A}_{GT}^{CA}$} &\multirow{2}{*}{$\tilde{A}_{GT}^{SSD}$}  &\multicolumn{2}{c}{IBM-2}  &ISM\footnotemark[1] &\multicolumn{2}{c}{exp\footnotemark[2]}
\\  \cline{4-5} \cline{7-8}
\T
&&&$^{CA}_{GT}$&$^{SSD}_{GT}$ & &$_{exp}^{CA}$ &$_{exp}^{SSD}$\\
\hline 
\T
48		&7.72\footnotemark[3] 	& 					&0.10 	&			&0.05 		&$0.038(3)$&\\
76		&9.41\footnotemark[3] 	& 								&0.24  &			&0.15 						&$0.118(5)$&\\
82		&10.1\footnotemark[3] 	& 								&0.18 	& 		&0.15 						&$0.083(4)$&\\
96		&11.0 					&2.20							&0.10		&0.51	& 							&$0.080(4)$&$0.075(4)$\\
100		&11.2 					&1.69							&0.13		&0.89	& 							&$0.206(7)$&$0.185(6)$\\
110		&11.8 					&1.89							&0.13		&0.80	& 							&&\\
116		&12.1 					&1.88				&0.10	&0.63		& 			&$0.114(5)$&$0.106(4)$\\
124		&12.5 					&									&0.12	&		& 							&&\\
128		&12.5\footnotemark[3] 	&								&0.15			&		& 					&$0.044(6)$&\\

130		&13.3\footnotemark[3] 	&								&0.13	&		&0.07 						&$0.031(4)$&\\
136		&13.1 					&								&0.11	&		&0.06 						&$0.0182(17)$&\\
148		&13.6 					&								&0.05	&		& 							&&\\
150		&13.7 					&1.88							&0.06		&0.42	& 							&$0.058(4)$&$0.052(4)$\\
154		&13.9 					&								&0.07&		& 							&&\\
160		&14.2 					&								&0.11&		& 							&&\\
198		&15.8 					&								&0.03&		& 							&&\\
\end{tabular}
\footnotetext[1]{Reference~\cite{caurier2007}.}
\footnotetext[2]{Reference~\cite{kotila}.}
\footnotetext[3]{Reference~\cite{haxton}.}
\end{table*}
\end{center}
\end{ruledtabular}

\begin{ruledtabular}
\begin{center}
\begin{table*}[cbt!]
\caption{\label{table14}Value of $g_{A,eff}$ extracted from experiment.
}
\begin{tabular}{ccccccc}
\multirow{3}{*}{Nucleus}&  \ensuremath{\tau_{1/2,\exp}(10^{18}} yr)\footnotemark[1] &\multicolumn{2}{c}{\ensuremath{\tau_{1/2}(10^{18}} yr)} &\multicolumn{2}{c}{\ensuremath{g_{A,eff}}} &\ensuremath{g_{A,eff}}\\ 
&exp &\multicolumn{2}{c}{IBM-2} 	&\multicolumn{2}{c}{IBM-2} 						&ISM\\ \cline{3-4} \cline{5-6}
\T
&&$^{CA}_{GT}$&$^{SSD}_{GT}$ &$^{CA}_{GT}$&$^{SSD}_{GT}$ &\\ \hline

\T
$^{48}$Ca	&$44^{+6}_{-5}$				&2.30	& 		&0.61(2)	&			&$0.90(3)$\\
$^{76}$Ge	&$1500\pm100$				&144	&		&0.71(1) 	&			&$0.90(2)$\\
$^{82}$Se	&$92\pm7$					&7.68	& 		&0.68(1) 	&			&$0.74(2)$\\
$^{96}$Zr	&$23\pm2$					&5.31	&0.187 	&0.88(2) 	&$0.38(1)$	&\\
$^{100}$Mo	&$7.1\pm0.4$				&6.46	&0.117 	&1.24(2)	&$0.46(1)$	&\\
$^{116}$Cd	&$28\pm2$					&14.5	&0.306	&1.08(2)	&$0.41(1)$	&\\
$^{128}$Te	&$1900000\pm400000$			&65600	&1170	&0.55(3) 	
&&\\
$^{130}$Te	&$680^{+120}_{-110}$		&15.5	&		&0.49(2)	&			&$0.67(3)$\\
$^{136}$Xe	&$2110\pm 250$\footnotemark[2]	&23.0	&	&0.41(2)	&			&$0.57(2)$\\
$^{150}$Nd	&$8.2\pm0.9$				&3.21	&0.048 &1.00(3)	&$0.35(1)$	&\\
\end{tabular}
\footnotetext[1]{Reference~\cite{barabash}.}
\footnotetext[2]{Reference~\cite{exo}.}
\end{table*}
\end{center}
\end{ruledtabular}

\begin{figure}[h]
\begin{center}
\includegraphics[width=8.6cm]{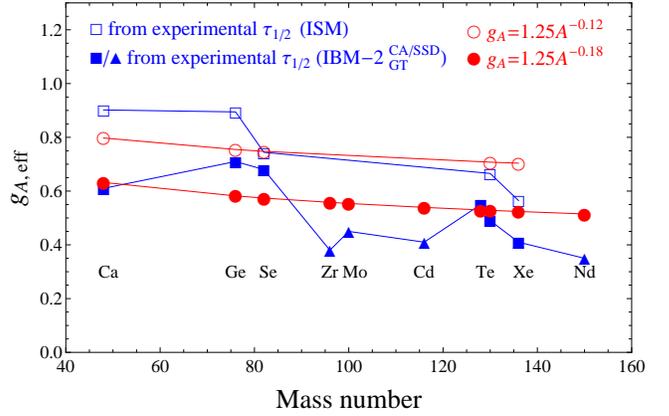} 
\end{center}
\caption{\label{fig11}(Color online) Value of $g_{A,eff}$ extracted from experiment for IBM-2 and the ISM.}
\end{figure}

If we write the matrix elements $M_{2\nu}$ as
\begin{equation}
M_{2\nu}^{eff}=\left(\frac{g_{A,eff}}{g_A} \right)^2 M_{2\nu},
\end{equation}
where $(g_{A, eff}/g_A)=q$ is a quenching factor, by comparing the experimental values $M_{2\nu,exp}$ with the calculated values (or the experimental half-lives with those calculated using PSFs of \cite{kotila}) we can extract the values of $g_{A, eff}$. These are given in Table~\ref{table14} and Fig.~\ref{fig11} for IBM-2 (GT) and the ISM. 
As mentioned in Sec.~\ref{sect0nu}, the renormalization of $g_A$ to $g_{A,eff}$ is due to two main reasons: (1) limitation of the model space in which calculations are done and (2) omission of non-nucleonic degrees of freedom ($\Delta, N^*$, ...). As a result, one expects $g_{A,eff}$ to have a smooth behavior with $A$ to which shell effects are superimposed. We see from Fig.~\ref{fig11} that this is approximately the case if we assume SSD in $_{40}$Zr, $_{42}$Mo, $_{48}$Cd, and $_{60}$Nd. This is consistent with previous analyses \cite{domin1,domin2}. The smooth behavior can be parametrized as $g_{A, eff}^{\rm{IBM-2}}=1.269A^{-\gamma}$, with  $\gamma=0.18$ for IBM-2 (GT). This gives for the neutron (A=1) the free value. The same type of analysis can be done for the ISM. The values of $g_{A, eff}$ extracted by comparing the calculated and experimental matrix elements are also shown in Table~\ref{table14} and Fig.~\ref{fig11}. We see that $g_{A, eff}$ in the ISM has the same behavior as in IBM-2, except for larger value. It can be parametrized as $g_{A, eff}^{\rm{ISM}}=1.269A^{-\gamma}$ with $\gamma=0.12$. In Ref.~\cite{caurier2007} the value 0.93 was used for $^{48}$Ca, $^{76}$Ge and $^{82}$Se and 0.71 for $^{130}$Te and $^{136}$Xe.

The question of how to extract $g_{A, eff}$ in the QRPA has been the subject of many investigations \cite{suhonen1998}. In this case $g_{A, eff}$ can be extracted either from $2\nu\beta\beta$ or from single-$\beta$ decay \cite{suhonen2005}. We do not discuss this extraction here but simply note that the values extracted are similar but larger than those in Table~\ref{table14} and Fig.~\ref{fig11}.

Values of $(g_{A, eff})^2$ can also be extracted from single-$\beta$ decay or electron capture using a Fermi-surface quasiparticle (FSQP) model \cite{ejiri09} where
\begin{equation}
(g_{A, eff})^2=g_i^{eff}g_f^{eff}
\end{equation}
is the product of $g_i^{eff}$ for the transition from even-even to odd-odd nuclei and $g_f^{eff}$ for the transition from odd-odd to even-even nuclei. The values obtained in this way \cite{ejiri10} are also similar to those in Table~\ref{table14} and Fig.~\ref{fig11}. Finally, very recently, values of $g_{A,eff,2\nu\beta\beta}^{\rm{IBM-2}}$ have been extracted from a $2\nu\beta\beta$ calculation without the closure approximation for $^{128,130}$Te$\rightarrow ^{128,130}$Xe decay with similar results \cite{yos12}. 
As one can see from the discussion in the paragraphs above, the extraction of the actual value of $g_{A,eff}$ is highly dependent on the model calculations and the assumptions  made. All extractions, however, indicate values of $g_{A,eff}$in the range $g_{A,eff}^{\rm{ISM}}\sim 0.57-0.90$ and $g_{A,eff}^{\rm{IBM-2}}\sim 0.35-0.71$ depending on mass number $A$  and on the SSD or CA approximation, with decreasing trend with $A$.

It is of considerable interest to analyze the impact that the quenching of $g_A$ to $g_{A,eff}$ observed in single-$\beta$ and $2\nu\beta\beta$ decay may have on $0\nu\beta\beta$. The question of whether or not the quenching of $g_A$ is the same in $2\nu\beta\beta$ as in $0\nu\beta\beta$ is the subject of debate, since only the states $1^+$ and $0^+$ in the intermediate odd-odd nucleus contribute to $2\nu\beta\beta$, while all multipoles contribute to $0\nu\beta\beta$. Two lines of thought have been considered : (1) Only GT ($1^+$) is quenched and other multipoles are not. (2) All multipoles are equally quenched. The experimental information on higher multipoles is meager, with only some hints coming from muon capture. The contribution of different intermediate states $J^{\pm}$ to $0\nu\beta\beta$ decay in $^{100}$Mo was investigated in Ref.~\cite{simkovic1} within the framework of QRPA-T\"u. It was found that the contribution of $1^+$ is sizeble, of opposite sign of that of the other multipoles, and very much parameter ($g_{pp}$) dependent. In view of this sizable contribution, even if the other multipoles are not quenched, there is going to be an effect coming from the $1^+$ multipole.

In order to investigate the possible impact of quenching of $g_A$, we present in Table~\ref{table15}, the predicted half-lives under the assumption of "maximal quenching" in which all multipoles are quenched with
\begin{equation}
\begin{split}
g_{A,eff}^{\rm{IBM-2}}&=1.269A^{-0.18},\\
g_{A,eff}^{\rm{ISM}}&=1.269A^{-0.12}.
\end{split}
\end{equation}
In the same table we show for comparison the unquenched values with $g_A=1.269$ discussed in Sec.~\ref{sect0nu}, and shown in Table~\ref{table10}. We observe that while the ISM unquenched values are a factor of approximately 2 smaller than IBM-2 values, the quenched values of both calculations are similar, since the smaller calculated matrix elements in the ISM are compensated in part by a larger value of $g_{A,eff}$. (This statement is not correct in $^{48}$Ca where the simple parametrization $1.269A^{-\gamma}$ fails.) It appears therefore that the difference of a factor of 2 in the calculated nuclear matrix elements in IBM-2 and the ISM  is simply due to the difference in the size of the model space and thus in the renormalization of $g_A$.

Another important question in this context is whether or not $g_V$ is quenched. From the conserved vector current (CVC) we expect $g_V$ not to be quenched, at least as far as the contribution of ($\Delta, N^*,...$) is concerned. On the other side, the size of the model space certainly affects the Fermi matrix elements, through the overlap of the initial and final wave functions and their isospin purity. Thus, if one defines
\begin{equation}
\left(\frac{g_{V,eff,\beta}}{g_V} \right)=\frac{\left|M_{exp,\beta} \right|}{\left|M_{th,\beta} \right|},
\end{equation}
where $g_V=1$, one may reasonably expect a quenching of the Fermi matrix elements as well. Whether or not the quenching factor for $g_{V,eff,\beta}$ is the same as for $g_{A,eff,\beta}$ is not clear. We are currently investigating this question within the context of IBM-2. Within this model, it appears also that the question of isospin violation can be dealt with by means of a quenching of the Fermi matrix elements. The question of how to project into states of good isospin was investigated years ago by introducing the concept of F-spin (isospin of the pairs) \cite[p.134]{iac11}. Because of the complexity of the problem, we do not discuss it here but defer it to a subsequent publication. In the columns "maximally quenched" in Table~\ref{table15} we have assumed equal quenching both for $g_V$ and $g_A$ and thus no quenching in the ratio $g_V/g_A$. This assumption introduces an additional error of about 10\% in the quenched calculation.

In conclusion, Table~\ref{table15} gives ranges of expected half-lives based on IBM-2 and the ISM calculations for unquenched, $g_A=1.269$, $g_V=1$, and "maximally quenched" values of $g_{A,eff}$, $g_{V,eff}$. The actual situation may in fact be in between these two extreme values. Similar analyses have been done within the QRPA except that a quenched value $g_{A,eff}=1.0$ is used while $g_{V,eff}=1$ is unquenched \cite{simkovic1}.

\begin{ruledtabular}
\begin{center}
\begin{table*}[cbt!]
\caption{\label{table15}Predicted half-lives in $0\nu\beta\beta$ decay with unquenched and maximally quenched $g_A$,  $g_{A,eff}^{\rm{IBM-2}}$ and $g_{A,eff}^{\rm{ISM}}$ obtained from $2\nu\beta\beta$ decay.}
\begin{tabular}{lcccc}
Decay&\multicolumn{4}{c}{\ensuremath{\tau_{1/2}^{0\nu}}(\ensuremath{10^{24}}yr)}\\ \cline{2-5}
\T &\multicolumn{2}{c}{IBM-2} &\multicolumn{2}{c}{ISM}\\ \cline{2-3} \cline{4-5}
\T
Decay  &unquenched &maximally quenched	&unquenched &maximally quenched\\
 \hline
 \T
$^{48}$Ca$\rightarrow ^{48}$Ti		&1.03	&16.8	&13.9	&89.2\\
$^{76}$Ge$\rightarrow ^{76}$Se		&1.45 	&32.8	&8.65	&69.1\\
$^{82}$Se$\rightarrow ^{82}$Kr		&0.52 	&12.4	&2.22	&18.5\\
$^{96}$Zr$\rightarrow ^{96}$Mo		&0.77	&20.5		&&\\
$^{100}$Mo$\rightarrow ^{100}$Ru	&0.46 	&12.5		&&\\
$^{110}$Pd$\rightarrow ^{110}$Cd	&1.60 	&47.1		&&\\
$^{116}$Cd$\rightarrow ^{116}$Sn 	&0.78	&24.0		&&\\
$^{124}$Sn$\rightarrow ^{124}$Te 	&0.91	&29.2	&2.73	&27.6\\
$^{128}$Te$\rightarrow ^{128}$Xe 	&8.53	&281	&33.5	&344.4\\
$^{130}$Te$\rightarrow ^{130}$Xe 	&0.44	&14.5	&1.70	&17.6\\
$^{136}$Xe$\rightarrow ^{136}$Ba 	&0.62	&21.5	&2.39	&25.3\\
$^{148}$Nd$\rightarrow ^{148}$Sm 	&2.54	&97.4		&&\\
$^{150}$Nd$\rightarrow ^{150}$Sm 	&0.30	&11.0		&&\\
$^{154}$Sm$\rightarrow ^{154}$Gd 	&5.34	&201		&&\\
$^{160}$Gd$\rightarrow ^{160}$Dy 	&0.80	&31.0		&&\\
$^{198}$Pt$\rightarrow ^{198}$Hg 	&3.77	&170		&&\\
\end{tabular}
\end{table*}
\end{center}
\end{ruledtabular}

\section{Conclusions}

In this article we have presented a consistent evaluation of nuclear
matrix elements in 0$\nu\beta\beta$ and 0$\nu_{h}\beta\beta$ decay, Sec.~\ref{sect0nu}, and 2$\nu\beta\beta$ 
decay, Sec.~\ref{sect2nu}, within the framework of IBM-2 in the closure approximation. All calculations can be done
simultaneously by replacing the neutrino potential
$v(p)$ as summarized in Appendix A. While the closure approximation  is expected to be good for $0\nu\beta\beta$ and $0\nu_h\beta\beta$ decay since the virtual neutrino momentum is of order 100 MeV/$c$ and thus much larger than the scale of nuclear excitations, it is not expected to be good for $2\nu\beta\beta$ decay where the neutrino momentum  is of order of few MeV/$c$ and thus of the same scale  of nuclear excitation. Furthermore, for $2\nu\beta\beta$, the single state dominance may be a better approximation. Hence the $2\nu\beta\beta$ calculation in Sec.~\ref{sect2nu} should be viewed only as an estimate.

By using the $0\nu\beta\beta$ matrix elements and phase-space factors of Ref.~\cite{kotila}, we have calculated the expected $0\nu\beta\beta$ half-lives in all nuclei of interest with $g_A=1.269$ and $g_V=1$, given in Table~\ref{table10} and Fig.~\ref{fig7}. This is the main result of this paper, and should be compared with other calculations, QRPA, ISM, DFT, and HFB, with the same (or similar) values of $g_A=1.25-1.269$ and $g_V=1$.

Finally, in Sec.~\ref{sect2nu}, we have examined the impact that a quenching of $g_A$ may have on $0\nu\beta\beta$ decay and reported in Table~\ref{table15} results of a quenched calculation with the quenching factor extracted from $2\nu\beta\beta$ decay. This calculation is speculative since we have no experimental information to confirm whether or not quenching is the same for all multipoles in the intermediate nucleus. Our assessment is that, while in the unquenched case the situation is such that current (GERDA and CUORE) and planned experiments may reach accuracies to detect at least the inverted hierarchy of Fig.~\ref{fig8}, in the quenched case only the degenerate case can be detected in the foreseeable future.
The results presented here point to the necessity of further studies and refinements, the crucial ones being: (i) an improved treatment of the Fermi matrix elements, (ii) an improved treatment of SRC, and (iii) the determination of the quenching factors $g_{A,eff}$  for all multipoles in $0\nu\beta\beta$ decay. The latter is of importance not only for IBM-2 but also for all other model calculations.

\section{Acknowledgements}

This work was supported in part by U.S.D.O.E. Grant DE-FG-02-91ER-40608
and Fondecyt Grant No. 1120462. We wish to thank all the experimental
groups that have stimulated our work, in particular A.\ Bettini,
S. E. Elliott, E. Fiorini, G. Gratta, A. McDonald, S. Schoenert, and
K. Zuber.

\section{Appendix A: Neutrino potentials and their radial integrals}

The neutrino potentials used in this article are given in Table~\ref{table16}.
The function $H(r)$ is the Fourier-Bessel transform of $2\pi^{-1}[p(p+\tilde{A})]^{-1}$,
and is given in \cite{tomoda}, Appendix 2, and in Eq.~(19) of \cite{simkovic}.
It does not have an explicit form. We note, however that, when the
closure energy $\tilde{A}$ goes to zero, then $v(p)=2\pi^{-1}p^{-2},$
and its Fourier-Bessel transform becomes the Coulomb potential, $H(r)\rightarrow 1/r$.
The neutrino potential is a long-range potential, since the mass of
the exchanged particle is very small. The situation is opposite in
the case of heavy neutrino exchange. In this case, the mass of the
exchanged particle is very large and thus the potential is a contact
interaction $\delta(r)/r^{2}$. For 2$\nu\beta\beta$ decay, the potential
does not have a radial dependence and thus it is a contact interaction
in momentum space. The values of $\tilde{A}$ used in this article
are given in Table~\ref{table13}. The radial integrals of the neutrino potential
are best calculated in momentum space using the Horie method \cite{horie}
as discussed in the Appendix A of \cite{barea}, with harmonic oscillator
single-particle wave functions with oscillator parameter $\nu=M\omega/\hslash$,
where $M$ is the nucleon mass. In this article, we take $\nu=\nu_{0}A^{-1/3}$,
where $A$ is the mass number and $\nu_{0}=0.994$~fm$^{-2}$.
\begin{table}[h]
\caption{\label{table16}Neutrino potentials used in this article.}
\begin{ruledtabular}
\begin{center}
 \begin{tabular}{ccc}
  Transition  &  \ensuremath{V(r)} &  \ensuremath{v(p)}\\
\hline
\T
 0\ensuremath{\nu\beta\beta} &  \ensuremath{H(r)} &  \ensuremath{\frac{2}{\pi}\frac{1}{p(p+\tilde{A})}}\\
0\ensuremath{\nu_{h}\beta\beta} &  \ensuremath{\frac{1}{m_{e}m_{p}}\frac{\delta(r)}{r^{2}}} &  \ensuremath{\frac{2}{\pi}\frac{1}{m_{e}m_{p}}}\\
2\ensuremath{\nu\beta\beta} &  \ensuremath{1} &  \ensuremath{\frac{\delta(p)}{p^{2}}}
\end{tabular}
\end{center}
\end{ruledtabular}
\end{table}

\section{Appendix B: Single-particle energies and strength of interaction}

In order to calculate the pair structure constants we need the single-particle
and single-hole energies and strength of interaction. We give in Tables~\ref{table17}-\ref{tab:spl82-126}, the single-particle and single-hole energies used in this article.
\begin{table}[cbt!]
 \caption{\label{table17}SDI strength values $A_{1}$ and single-particle and single-hole energies
(in MeV) in the $N,Z=28-50$ shell. The energies are taken
from the spectra of $^{57}\mbox{Cu}$ for proton particles, from isotones
$N=50$ for proton holes, and from the spectra of $^{57}\mbox{Ni}$
for neutron holes.}
\label{tab:spl28-50}\vspace*{0.5cm}
\begin{ruledtabular}
\begin{tabular}{cccc}
Orbital  & %
\begin{tabular}{c}
Protons\tabularnewline
(particles)\tabularnewline
$A_{1}=0.366$\tabularnewline
\end{tabular} & %
\begin{tabular}{c}
Protons\tabularnewline
(holes)\tabularnewline
$A_{1}=0.264$\tabularnewline
\end{tabular} & %
\begin{tabular}{c}
Neutrons\tabularnewline
(holes)\tabularnewline
$A_{1}=0.280$\tabularnewline
\end{tabular}\tabularnewline
\hline 
\T
$2p_{1/2}$  & 1.106  & 0.931  & 1.896\tabularnewline
$2p_{3/2}$  & 0.000  & 2.198  & 3.009\tabularnewline
$1f_{5/2}$  & 1.028  & 2.684  & 2.240\tabularnewline
$1g_{9/2}$  & 3.009  & 0.000  & 0.000\tabularnewline
\end{tabular}
 \end{ruledtabular} 
\end{table}
\begin{table}[cbt!]
 \caption{SDI strength values $A_{1}$ and single-particle and single-hole energies
(in MeV) in the $N,Z=50-82$ shell. The energies are taken
from the spectra of $^{133}\mbox{Sb}$ for protons particles, from
the spectra of $^{207}\mbox{Tl}$ for proton holes, from the spectra
of $^{91}\mbox{Zr}$ for neutron particles, and from the spectra of $^{131}\mbox{Sn}$
for neutron holes.}
\label{tab:spl50-82}
\begin{ruledtabular}
\begin{centering}
\begin{tabular}{ccccc}
Orbital  & %
\begin{tabular}{c}
Protons\tabularnewline
(particles)\tabularnewline
$A_{1}=0.221$\tabularnewline
\end{tabular} & %
\begin{tabular}{c}
Protons\tabularnewline
(holes)\tabularnewline
$A_{1}=0.200$\tabularnewline
\end{tabular} & %
\begin{tabular}{c}
Neutrons\tabularnewline
(particles)\tabularnewline
$A_{1}=0.269$\tabularnewline
\end{tabular} & %
\begin{tabular}{c}
Neutrons\tabularnewline
(holes)\tabularnewline
$A_{1}=0.163$\tabularnewline
\end{tabular}\tabularnewline
\hline 
\T
$3s_{1/2}$  & 2.990  & 0.000  & 1.205  & 0.332\tabularnewline
$2d_{3/2}$  & 2.690  & 0.350  & 2.042  & 0.000\tabularnewline
$2d_{5/2}$  & 0.960  & 1.670  & 0.000  & 1.655\tabularnewline
$1g_{7/2}$  & 0.000  & 2.700  & 2.200  & 2.434\tabularnewline
$1h_{11/2}$  & 2.760  & 1.340  & 2.170  & 0.070\tabularnewline
\end{tabular}
\par\end{centering}
 \end{ruledtabular} 
\end{table}
\begin{table}[cbt!]
 \caption{SDI strength values $A_{1}$ and single-particle energies (in MeV)
in the $N=82-126$ shell. The energies are taken from \protect\cite{pittel}
for neutron particles and from the spectra of $^{208}\mbox{Pb}$ for
neutron holes.}
\label{tab:spl82-126}
\begin{ruledtabular}
\begin{tabular}{ccc}
Orbital  & %
\begin{tabular}{c}
Neutrons \tabularnewline
(particles)\tabularnewline
$A_{1}=$0.147\tabularnewline
\end{tabular} & %
\begin{tabular}{c}
Neutrons\tabularnewline
(holes)\tabularnewline
$A_{1}=$0.150\tabularnewline
\end{tabular}\tabularnewline
\hline 
\T
$3p_{1/2}$  & 2.250  & 0.000\tabularnewline
$3p_{3/2}$  & 1.500  & 0.900\tabularnewline
$2f_{5/2}$  & 2.600  & 0.570\tabularnewline
$2f_{7/2}$  & 0.000  & 2.340\tabularnewline
$1h_{9/2}$  & 2.450  & 3.410\tabularnewline
$1i_{13/2}$  & 2.800  & 1.630\tabularnewline
\end{tabular}
\end{ruledtabular} 
\end{table}
We generate the pair structure constants by diagonalizing the surface
delta interaction (SDI) in the two identical particle states, $pp$ and
$nn$. The strength of the (isovector) interaction, $A_{1}$, is also
given in Tables~\ref{table17}-\ref{tab:spl82-126}. It is obtained by fitting the 2$^{+}$ 0$^{+}$
energy difference in nuclei with either two protons (proton holes)
or two neutrons (neutron holes). 
For $^{48}$Ca$\rightarrow$ $^{48}$Ti decay, we need also the strength
of the interaction in the $1f_{7/2}$ shell, given by $A_{1}=0.510$~MeV.
The calculation of the pair structure constants can be improved by
a better choice of the interaction and of the single-particle energies.
We have tried different choices of the single-particle energies and
included the variation of the corresponding radial integrals in the
estimate of the sensitivity to parameter changes.

\section{Appendix C: Parameters of the IBM-2 Hamiltonian}

A detailed description of the IBM-2 Hamiltonian is given in \cite{iac1} and \cite{otsukacode}. For most nuclei, the Hamiltonian parameters are taken from the literature \cite{Duval83, Kaup83, Kaup79, Shlomo92, Isacker80, Kim96, Giannatiempo91, Sambataro82, Iachello96, Puddu80, ScholtenPhD, Bijker80, Barfield83}. The values of the Hamiltonian parameters, as well as the references from which they were taken, are given in Table~\ref{tab:ibm2parameters}. The quality of the description can be seen from these references and ranges from very good to excellent (see Fig.~\ref{spectra}). The only nuclei for which we have done new calculations are $^{48}$Ti, $^{96}$Zr, $^{124}$Sn, $^{136}$Xe , $^{160}$Gd,  and $^{160}$Dy. The new calculations are done using the program NPBOS \cite{otsukacode} adapted by J. Kotila. They include energies, $B(E2)$ values, quadrupole moments, $B(M1)$ values, magnetic moments, etc. The calculations for $^{160}$Gd and $^{160}$Dy have just been published \cite{kot12b}. A paper with those for $^{48}$Ti is in preparation \cite{kot12c}. The quality of these, as well as of the unpublished results for $^{96}$Zr is equal to that of the results obtained previously \cite{Duval83, Kaup83, Kaup79, Shlomo92, Isacker80, Kim96, Giannatiempo91, Sambataro82, Iachello96, Puddu80, ScholtenPhD, Bijker80, Barfield83}. For the semi-magic nuclei $^{116-124}$Sn and $^{136}$Xe, we have obtained the parameters by a fit to the energy of the low-lying states using the same procedure as in Ref.~\cite{Iachello96} for $^{116}$Sn, while $^{48}$Ca has been taken as doubly magic. This procedure is compatible with the generalized seniority (GS) scheme, which appears to be good for semi-magic nuclei,
 as extensively discussed in the 1980s for pairing plus quadrupole interactions, and as shown recently for realistic interactions \cite{cap12}.

\begin{table*}[ctb!]
 \caption{Hamiltonian parameters employed in the IBM-2 calculation of the final
wave functions along with their references.}
\label{tab:ibm2parameters}
\begin{ruledtabular}
\begin{centering}
\begin{tabular}{cccccccccccccccccccc}
Nucleus  & $\epsilon_{d_{\nu}}$  & $\epsilon_{d_{\pi}}$  & $\kappa$  & $\chi_{\nu}$  & $\chi_{\pi}$  & $\xi_{1}$  & $\xi_{2}$  & $\xi_{3}$  & $c_{\nu}^{(0)}$  & $c_{\nu}^{(2)}$  & $c_{\nu}^{(4)}$  & $c_{\pi}^{(0)}$  & $c_{\pi}^{(2)}$  & $c_{\pi}^{(4)}$  & $\omega_{\nu\nu}$  & $\omega_{\pi\pi}$  & $\omega_{\nu\pi}$  & $w_{\nu}$  & $y_{\nu}$\tabularnewline
\hline 
\T
$^{48}\mbox{Ti}$\footnotemark[1]  & 1.11  & 1.11  & -0.20  & -0.30  & -0.70  & 1.00  & 1.00  & 1.00  &  &  &  &  &  &  &  &  &  &  & \tabularnewline
$^{76}\mbox{Ge}$ \cite{Duval83}  & 1.20  & 1.20  & -0.21  & 1.00  & -1.20  & -0.05  & 0.10  & -0.05  &  &  &  &  &  &  &  &  &  &  & \tabularnewline
$^{76}\mbox{Se}$ \cite{Kaup83}  & 0.96  & 0.96  & -0.16  & 0.50  & -0.90  &  &  & -0.10  &  &  &  &  &  &  &  &  &  &  & \tabularnewline
$^{82}\mbox{Se}$ \cite{Kaup83}  & 1.00  & 1.00  & -0.28  & 1.14  & -0.90  &  &  & -0.10  &  &  &  &  &  &  &  &  &  &  & \tabularnewline
$^{82}\mbox{Kr}$ \cite{Kaup79}  & 1.15  & 1.15  & -0.19  & 0.93  & -1.13  & -0.10  &  & -0.10  &  &  &  &  &  &  &  &  &  &  & \tabularnewline
$^{96}\mbox{Zr}$\footnotemark[1]  & 1.00  & 1.00  & -0.20  & -2.20  & 0.65  &  &  &  &  &  &  &  &  &  & 0.17  & 0.17  & 0.33  &  & \tabularnewline
$^{96}\mbox{Mo}$ \cite{Shlomo92}  & 0.73  & 1.10  & -0.09  & -1.20  & 0.40  & -0.10  & 0.10  & -0.10  & -0.50  & 0.10  &  &  &  &  &  &  &  &  & \tabularnewline
$^{100}\mbox{Mo}$ \cite{Shlomo92}  & 0.55  & 1.00  & -0.06  & -1.20  & 0.40  & -0.10  & 0.10  & -0.10  & -0.60  & 0.20  & 0.10  &  &  &  &  &  &  &  & \tabularnewline
$^{100}\mbox{Ru}$ \cite{Isacker80}  & 0.89  & 0.89  & -0.18  & -1.00  & 0.40  &  &  &  & 0.60  & 0.09  & -0.13  &  &  &  &  &  &  &  & \tabularnewline
$^{110}\mbox{Pd}$ \cite{Kim96}  & 0.78  & 0.60  & -0.13  & 0.00  & -0.30  & 0.20  & 0.04  & 0.00  & -0.26  & -0.29  & -0.30  & -0.26  & -0.29  & -0.03  &  &  &  &  & \tabularnewline
$^{110}\mbox{Cd}$ \cite{Giannatiempo91}  & 0.92  & 0.92  & -0.15  & -1.10  & -0.80  & 1.10  & 0.109  & 1.10  & 0.07  & -0.17  & 0.16  &  &  &  &  &  &  &  & \tabularnewline
$^{116}\mbox{Cd}$ \cite{Sambataro82}  & 0.85  & 0.85  & -0.27  & -0.58  & 0.00  & -0.18  & 0.24  & -0.18  & -0.15  & -0.06  &  &  &  &  &  &  &  &  & \tabularnewline
$^{116}\mbox{Sn}$ \cite{Iachello96}  & 1.32  &  &  &  &  &  &  &  & -0.50  & -0.22  & -0.07  &  &  &  &  &  &  & -0.06  & 0.04\tabularnewline
$^{124}\mbox{Sn}$\footnotemark[2]  & 1.10  &  &  &  &  &  &  &  & -0.30  & -0.16  & -0.20  &  &  &  &  &  &  & 0.30  & 0.02\tabularnewline
$^{124}\mbox{Te}$ \cite{Sambataro82}  & 0.82  & 0.82  & -0.15  & 0.00  & -1.20  & -0.18  & 0.24  & -0.18  & 0.10  &  &  &  &  &  &  &  &  &  & \tabularnewline
$^{128}\mbox{Te}$ \cite{Sambataro82}  & 0.93  & 0.93  & -0.17  & 0.50  & -1.20  & -0.18  & 0.24  & -0.18  & 0.30  & 0.22  &  &  &  &  &  &  &  &  & \tabularnewline
$^{128}\mbox{Xe}$ \cite{Puddu80}  & 0.70  & 0.70  & -0.17  & 0.33  & -0.80  & -0.18  & 0.24  & -0.18  & 0.30  &  &  &  &  &  &  &  &  &  & \tabularnewline
$^{130}\mbox{Te}$ \cite{Sambataro82}  & 1.05  & 1.05  & -0.20  & 0.90  & -1.20  & -0.18  & 0.24  & -0.18  & 0.30  & 0.22  &  &  &  &  &  &  &  &  & \tabularnewline
$^{130}\mbox{Xe}$ \cite{Puddu80}  & 0.76  & 0.76  & -0.19  & 0.50  & -0.80  & -0.18  & 0.24  & -0.18  & 0.30  & 0.22  &  &  &  &  &  &  &  &  & \tabularnewline
$^{136}\mbox{Xe}$\footnotemark[2]  &  & 1.31  &  &  &  &  &  &  &  &  &  & -0.04  & 0.01  & -0.02  &  &  &  &  & \tabularnewline
$^{136}\mbox{Ba}$ \cite{Puddu80}  & 1.03  & 1.03  & -0.23  & 1.00  & -0.90  & -0.18  & 0.24  & -0.18  & 0.30  & 0.10  &  &  &  &  &  &  &  &  & \tabularnewline
$^{148}\mbox{Nd}$ \cite{ScholtenPhD}  & 0.70  & 0.70  & -0.10  & -0.80  & -1.20  & -0.12  & 0.24  & 0.90  &  &  &  & 0.40  & 0.20  &  &  &  &  &  & \tabularnewline
$^{148}\mbox{Sm}$ \cite{ScholtenPhD}  & 0.95  & 0.95  & -0.12  & 0.00  & -1.30  & -0.12  & 0.24  & 0.90  &  &  &  &  & 0.05  &  &  &  &  &  & \tabularnewline
$^{150}\mbox{Nd}$ \cite{ScholtenPhD}  & 0.47  & 0.47  & -0.07  & -1.00  & -1.20  & -0.12  & 0.24  & 0.90  &  &  &  & 0.40  & 0.20  &  &  &  &  &  & \tabularnewline
$^{150}\mbox{Sm}$ \cite{ScholtenPhD}  & 0.70  & 0.70  & -0.08  & -0.80  & -1.30  & -0.12  & 0.24  & 0.90  &  &  &  &  & 0.05  &  &  &  &  &  & \tabularnewline
$^{154}\mbox{Sm}$ \cite{ScholtenPhD}  & 0.43  & 0.43  & -0.08  & -1.10  & -1.30  & -0.12  & 0.24  & 0.90  &  &  &  &  & 0.05  &  &  &  &  &  & \tabularnewline
$^{154}\mbox{Gd}$ \cite{ScholtenPhD}  & 0.55  & 0.55  & -0.08  & -1.00  & -1.00  & -0.12  & 0.24  & 0.90  &  &  &  & -0.20  & -0.10  &  &  &  &  &  & \tabularnewline
$^{160}\mbox{Gd}$ \cite{kot12b}  & 0.42  & 0.42  & -0.05  & -0.80  & -1.00  & 0.08  & 0.08  & 0.08  &  &  &  & -0.20  & -0.10  &  &  &  &  &  & \tabularnewline
$^{160}\mbox{Dy}$ \cite{kot12b}  & 0.44  & 0.44  & -0.06  & -0.80  & -0.90  & 0.08  & 0.08  & 0.08  &  &  &  & -0.05  & -0.15  &  &  &  &  &  & \tabularnewline
$^{198}\mbox{Pt}$ \cite{Bijker80}  & 0.58  & 0.58  & -0.18  & 1.05  & -0.80  & -0.10  & 0.08  & -0.10  & 0.00  & 0.02  & 0.00  &  &  &  &  &  &  &  & \tabularnewline
$^{198}\mbox{Hg}$ \cite{Barfield83}  & 0.55  & 0.55  & -0.21  & 1.00  & -0.40  &  & 0.08  &  & 0.37  & 0.25  & 0.16  &  &  &  &  &  &  &  & \tabularnewline
\end{tabular}
\footnotetext[1]{Parameters fitted to reproduce the spectroscopic
data of the low-lying energy states.}
\footnotetext[2]{GS parameters fitted to reproduce the spectroscopic data of the low-lying energy states.}
\par\end{centering}
 \end{ruledtabular} 
\end{table*}


\end{document}